\documentclass[aps,prd,reprint,amsmath,amssymb,superscriptaddress,floatfix]{revtex4}
\usepackage{graphicx}
\usepackage{amsfonts}
\usepackage{amsmath}

\usepackage{amssymb}
\usepackage{rotating}
\usepackage{booktabs}
\usepackage{xcolor}
\usepackage{soul}
\usepackage{color}
\usepackage{slashed}
\usepackage{multirow}
\usepackage{makecell}
\usepackage{epsf}
\usepackage{ulem}
\usepackage{cancel}
\usepackage{color,bm}
\usepackage{bm}
\usepackage{subfigure}
\usepackage{caption}
\usepackage{float}
\usepackage{mathtools}

\usepackage[colorinlistoftodos]{todonotes}

\usepackage[colorlinks=true,citecolor=cyan,urlcolor=blue,bookmarks=true,bookmarks=true,bookmarksopen=true,bookmarksnumbered=true,bookmarksopenlevel=3]{hyperref}

\usepackage[compat=1.1.0]{tikz-feynman}

\definecolor{airforceblue}{rgb}{0.36, 0.54, 0.66}
\definecolor{steelblue}{rgb}{0.27, 0.51, 0.71}
\definecolor{amber}{rgb}{1.0, 0.49, 0.0}

\allowdisplaybreaks

\begin{document}

% Title and author
\title{An Attention Based Neural Network for Jet Tagging}
\author{Jing Li}
\email{ lj948756370@mail.dlut.edu.cn }
\author{Hao Sun}
\email{ haosun@dlut.edu.cn }
%\footnote{lj948756370@mail.dlut.edu.cn \hspace{0.2cm} haosun@dlut.edu.cn}}
\affiliation{ Institute of Theoretical Physics, School of Physics, Dalian University of Technology, \\ No.2 Linggong Road, Dalian, Liaoning, 116024, P.R.China }
\date{\today}

% Abstract
\begin{abstract}

Convolutional neural networks are basic structures using jet images as input for the jet tagging problems. However, what they have learned during the training process is always difficult to understand just through feature maps. Inspired by the attention mechanism popular in machine learning fields, we propose a novel attention-based neural network (ABNN) to get insight of this problem. The ABNN combines a jet image with average jet images from the signal and the background to generate attention maps which show clearly the relevant importance according to the different origination of jets. Compared with networks in the similar architecture, this network achieves better performance, which indicates the potential of attention mechanism to use in other works. 

\vspace{0.5cm}
\end{abstract}

\maketitle
\setcounter{footnote}{0}

\section{INTRODUCTION}
\label{I}

After a collision of a bunch of particles, more particles are generated through the parton shower and the hadronization. Then they reach various detectors, e.g., calorimeters, to generate energy deposits to be observed by researchers. A jet is essentially one spray of particles observed. As the energy goes higher, highly boosted resonance will make this spray narrower and force the particles to be more collinear, which, as a result, generates the "fat jet". Using the rich substructure of such fat jets to analyze backwards the heavy hadron process plays an important role in the analysis on the Large Hadron Collider (LHC). A fundamental challenge is to distinguish the jets initiated by the desired process from the general and overwhelming QCD jet, which is called "jet tagging".

Over the past several years, many works apply the modern neural networks into this problem. See Refs.\cite{Kogler2018, Marzani2019, Kasieczka2019, Larkoski2017} for in depth reviews. Different jet representations come out for different machine learning models, such as images \cite{Almeida2015, Cogan2014, Baldi2016, Oliveira2015, Kasieczka2017, Komiske2016, Lin2018, Macaluso2018, Fraser2018, Moore2018, Diefenbacher2019, Chen2019}, sequences\cite{Guest:2016iqz, Pearkes:2017hku, Egan:2017ojy, Butter:2017cot, Erdmann:2018shi, Kasieczka:2018lwf}, graphs\cite{Abdughani2018}, and sets\cite{Komiske2018, Qu2019}. A visually straightforward and the first-used representation of jets in deep learning is the jet image. The energy depositions are used as pixel intensities in the pseudo-rapidity vs azimuthal plane. Other information like jet charge can be added to different channels of the jet image. Taking the jet image as input, many convolutional neural networks (CNNs) have been built to explore the potential of machine learning methods.

\cite{Oliveira2015} firstly introduces the deep neural networks to jet tagging. The convolutional architecture contains three sequential units (Conv + Max-Pool + Dropout), a local response normalization (LRN) layer, and two fully connected layers. \cite{Komiske2016} supplements the jet image with colors. The three parts: the transverse momentum in charged particles, transverse momentum in neutral particles, and pixel-level charged particle counts, construct the jet image. The network consists of three sets of a convolutional layer and max-pooling and a dense layer is followed. \cite{Lin2018} proposes a novel two-stream CNN. One stream acts on the full event information and the other one acts on the image of Higgs candidate jet. Each stream is followed by a dense layer and then they are connected to the final output neuron. \cite{Macaluso2018} introduces a number of improvements to the DeepTop tagger proposed by \cite{Kasieczka2017}. They augmented the tagger with more feature maps and more nodes on dense layers. Their CNN contains 6 convolutional layers and 3 dense layers. \cite{Chen2019} proposed 'CNN2', which has an asymmetric design for the $p_T$ channel and the jet charge $\mathcal{Q}_k$. One channel has 8 convolutional layers and 2 dense layers and the number of these layers of the other one are 5 and 1. The output of the two channels are concatenated to connect to the final dense layer.

Understanding what happens during the training process of CNNs is difficult. \cite{Oliveira2015, Lin2018} have tried using the feature maps to address this problem. But several feature maps do so less to that because of characteristics of CNN: each layer is connected to each other and the impact of intermediate feature maps on the initial image or the final results is obscure. We try to use the attention mechanism \cite{Vaswani2017, Jetley2018, Bello2019}    to bring some insights into this problem. The attention mechanisms calculate how much attention one network should pay to different regions of a jet image. The regions with higher attention are clearly expected to have more important features. Other regions, however, are suppressed by the low attention due to the potentially irrelevant and confusing information. Inspired by this, we propose the attention-based neural networks (ABNNs). The ABNNs generate the attention of a jet image by combining both signal and background average jet images.

For the fat jets initiated by heavy particles (e.g., top quarks and W, Z, and Higgs bosons), the average jet images show distinct multi-prong structures. However, for the general QCD jets, the average images do not show this characteristics. Based upon this point, the average jet images potentially have all features to instruct the neural network to focus on patches of one jet image.

The ABNN is used to tag jets from boosted Z bosons against QCD jets. The experiments show that it achieves better performance over networks in the similar architecure. By visualizing the attention of jet image samples, we show clearly which regions attract the most attention of the network. Under the designing idea of ABNNs, the attention from average jet images from both the signal and the background tends to become unified. Average jet images consisting of different numbers of samples are also to be used to evaluate the performance of ABNNs. The results show under the enhancement of the attention mechanism, the network become more robust than the one without it. 

The rest of the paper is organized as follows. In Sec.II, we introduce the attention mechanism and describe the ABNN architectures. Details in event generation configuration are in Sec.III. All results are shown and analyzed in Sec.IV. In Sec.V, with the benefit of the attention mechanism, we show the focus of ABNNs during the training process. We conclude our work and outlook in the last Sec.VI.

\section{ATTENTION-BASED NEURAL NETWORK}
\label{II}

\subsection{Attention mechanism}
\label{II.A}

In \cite{Jetley2018}, the attention map is used to visualize and interprete the inner reasoning process of CNNs, which is defined as: a scalar matrix representing the relative importance of layer activations at different 2D spatial locations with respect to the target task. The attention map is also called 'attention' for short. The way of introducing and calculating such attention map is the attention mechanism. According to the different schemes used to implement attention, it can be classified as post hoc network analysis and trainable attention mechanism. The post hoc method extract attention from already trained CNNs, while the latter does have an influence on the training process. There are two ways of the trainable attention mechanism: hard and soft attention. Hard attention is more like an image cropping method, i.e. only some patches are considered to be trained and other regions are left out. On the contrary, for the soft attention, all pixels are given different weights which reflect their relative importance. In this paper, we take the soft attention as the basic building block of the ABNNs. 

In high energy physics, we want to reconstruct events closer to the true scattering to get a greater signal significance. Much attention should be paid to the final states from a hard scattering event. Different jet grooming methods \cite{Butterworth2008, Ellis2009, Carrazza2019, Krohn2009a, Dasgupta2013, Larkoski2014, Dreyer2018} can be thought of as one kind of attention mechanism that improve the results of reconstruction. Here's another example of attention mechanism. When finding a person among many people in a photo, we use the memory of that person as a query to compare with all people shown. All the people can be seen as a number of key-value pairs. The key can be the features one has like the haircut, and the value corresponds to the pixel intensities of that person. The attention of the query and keys, which describe how close they are, can be calculated. The higher the score is, the more likely the person is the one we want to find. In general, the input we feed into a neural network can be seen as a key-value pair as well. The target task is the query. By defining a kind of compatibility measure between the key and the query, the attention mechanism can be added to standard stacked CNNs easily. In our work, the values weighted by the normalized compatibility scores are defined as attention.

Based on jet images to address the jet tagging task, different CNNs have been explored. The common way to show what kind of information the CNN has learned is to visualize several feature maps extracted from the intermediate layers. However, it's still hard to understand them since feature maps are connected one layer by one layer. When they are considered as a whole, they work out like a feature extractor to distill the useful information. If just concerned for some of them, their impact on the final results are difficult to figure out. Compared with them, the attention obtained could show the relative importance of each part of a jet image. Using all attention to connect directly to the final classifying layer, the focus of the whole model is shown clearly, which brings new insights into the inference process.

There are various ways to define the attention. In \cite{Vaswani2017}, scaled dot product attention has been used:
\begin{align}
    \text { Attention }(Q, K, V)=\operatorname{softmax}\left(\frac{Q K^{T}}{\sqrt{d_{k}}}\right) V\ .
\end{align}
The dot product of queries ($Q$) and keys ($K$) is scaled by the dimension of the key ($d_k$) to generate the compatibility scores, where $T$ means the transverse operation. The attention is obtained by weighting the values ($V$) with the compatibility scores normalized by the softmax function. In \cite{Jetley2018}, a parameterized attention is used:
\begin{align}
    \text { Attention }(Q, K, V)=\operatorname{softmax}\left((Q+K)W\right) V \ \ .
\end{align}
The keys and values are local feature maps extracted from intermediate layers and the queries are the global feature maps normally fed to the final linear layer. $W$ is a learnable weight matrix. In this paper, we use the parameterized attention.

\begin{figure}[htp]
    \centering
    \subfigure[]{
        \includegraphics[scale=0.5]{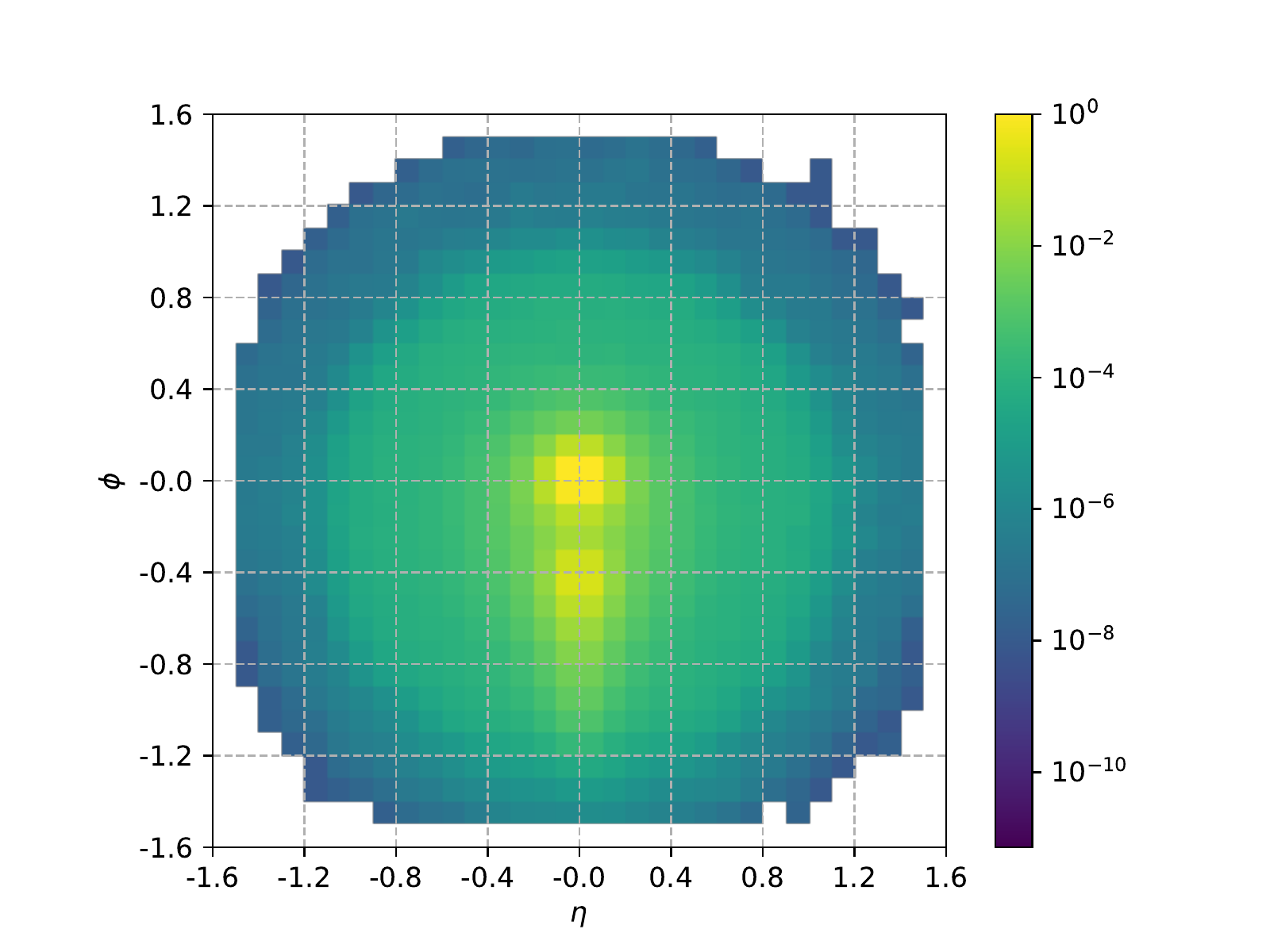}
    }
    \subfigure[]{
        \includegraphics[scale=0.5]{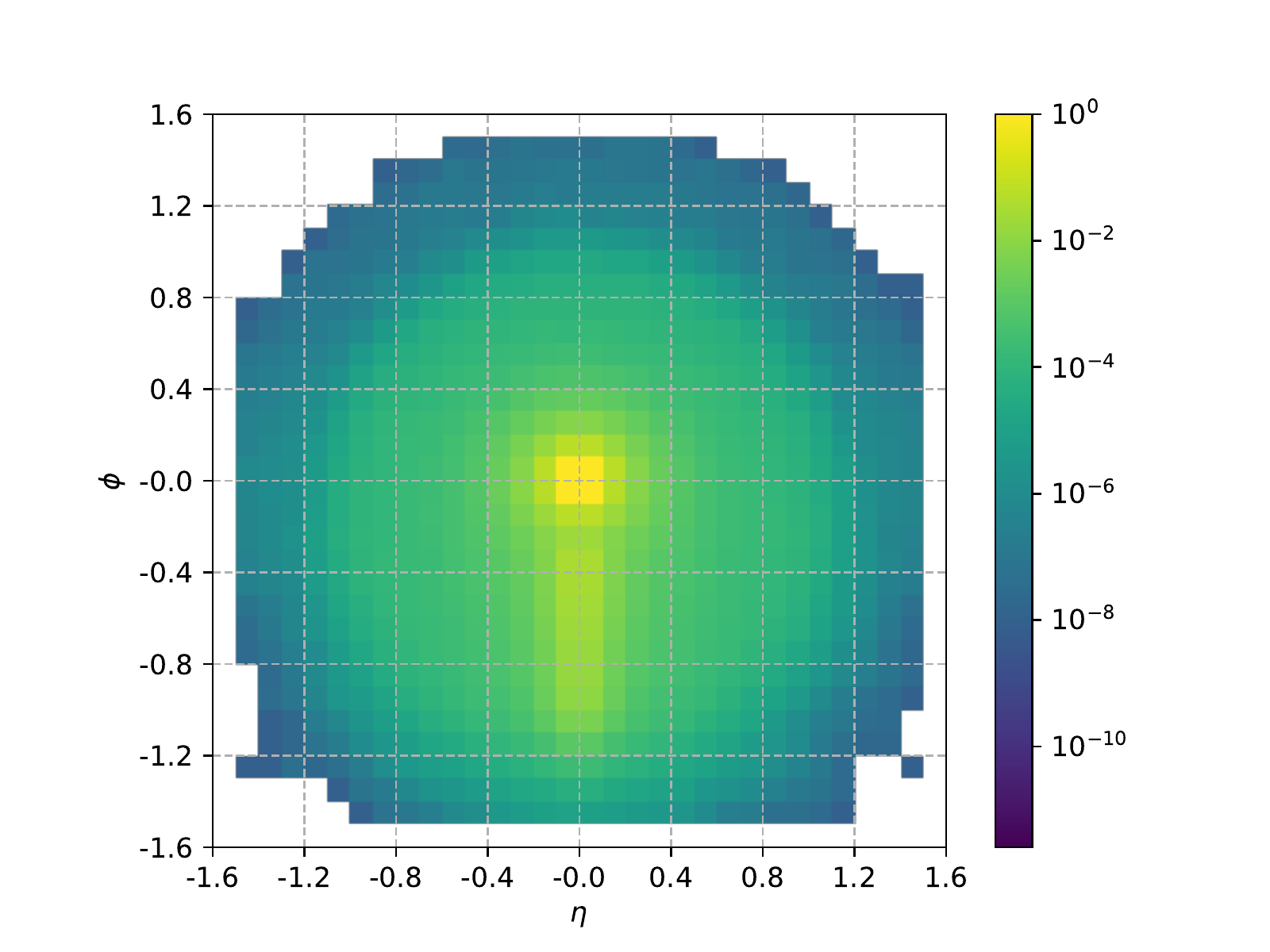}
    }
    \caption{Average jet images. The left is from the signal and the right is from the bakground.}
    \label{fig:Average jet images}
\end{figure}

In jet tagging tasks based on jet images, the input image itself is viewed as both a key and a value. We notice that the average jet images are a display of the entire jet images. They respectively show the essential characteristics of the signal and background, as shown in Fig.\ref{fig:Average jet images}.
The multi-prong feature obviously appears on the signal average jet image, while the image from the background has no such clear feature. Similar with the example in the photo above, these features are an important reference for us to find the signal out of the background. In addition to the potential features that the input jet image itself has, we regard the average jet image as a global feature, hoping to use it as a query to guide the neural network to notice the important parts of the input.

\subsection{Network architecture}
\label{II.B}

This section describes the attention-based neural networks (ABNNs). Before describing in detail, we give some necessary symbolic meanings. ConvBlock consists of a convolutional layer and a batch normalization layer, and outputs through the ReLU activation function. In order to deepen our network, the kernel size of the convolutional layer is set to 3x3, and the padding number is set to 1. This setting keeps the size of the input unchanged through the convolutional layer. AttnBlock calculates the corresponding normalized compatibility score and attention by key-value pairs and queries. In order to show the effect of the attention mechanism, we set up LinearBlock. The difference between it and AttnBlock is that there is no learnable parameter $W$, which means the output is obtained through an addition operation and a softmax function.

\begin{table}[]
\begin{tabular}{|c|c|c|c|}
\hline
            & FE1 & FE2 & FE3 \\ \hline
ConvBlock1  & $1 \to 8$         & $1 \to 8$         & $1 \to 8$         \\
ConvBlock2  &                   & $8 \to 8$         & $8 \to 8$         \\
ConvBlock3  &                   &                   & $8 \to 8$         \\
Max Pooling &                   &                   &                   \\
ConvBlock4  & $8 \to 16$        & $8 \to 16$        & $8 \to 16$        \\
ConvBlock5  &                   & $16 \to 16$       & $16 \to 16$       \\
ConvBlock6  &                   &                   & $16 \to 16$       \\
Max Pooling &                   &                   &                   \\
ConvBlock7  & $16 \to 32$       & $16 \to 32$       & $16 \to 32$       \\
ConvBlock8  &                   & $32 \to 32$       & $32 \to 32$       \\
ConvBlock9  &                   &                   & $32 \to 32$       \\
Max Pooling &                   &                   &                   \\
ConvBlock10 & $32 \to 64$       & $32 \to 64$       & $32 \to 64$       \\
Dropout     &                   &                   &                   \\\hline
\end{tabular}
\caption{Structures and channel variation of different feature extractors.}
\label{table:feature extractors}
\end{table}

Usually the input will be multiplied by parameter matrices and converted into the corresponding key-value pair and the query. Unlike this approach, we regard the standard stacked CNN as a kind of feature extractors (FEs). Feature maps of different dimensions in the intermediate stage will be extracted as key-value pairs and queries for the next step of calculation. In order to control the number of parameters, the maximum channel number of all FEs is fixed at 64. We extract more complex feature maps by increasing the number of ConvBlock with the same number of channels. The three FE structures are listed in Table \ref{table:feature extractors}.

\begin{figure}[htp]
    \centering
    \subfigure[ABNNs]{
        \includegraphics[scale=0.8]{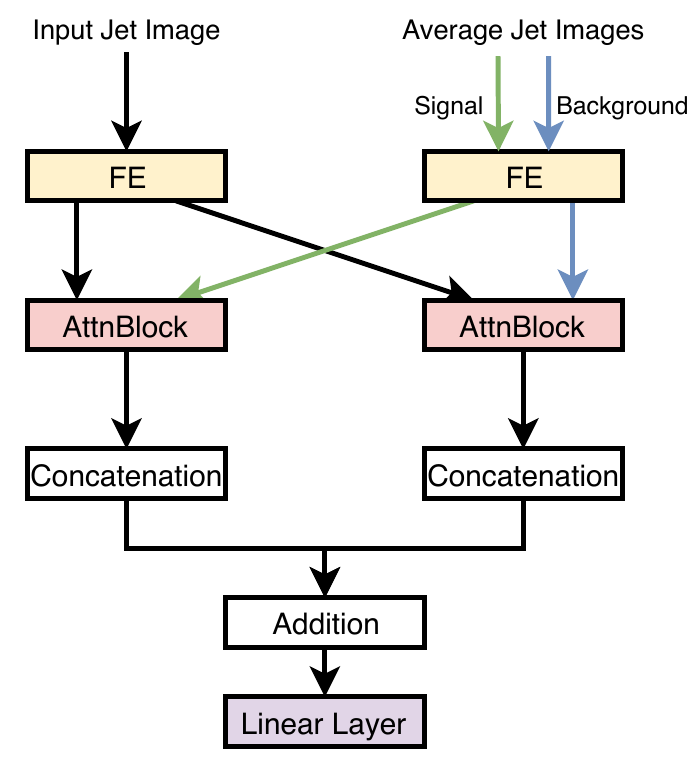}
        \label{ABNNs}}
    \subfigure[CNN-1]{
        \includegraphics[scale=0.8]{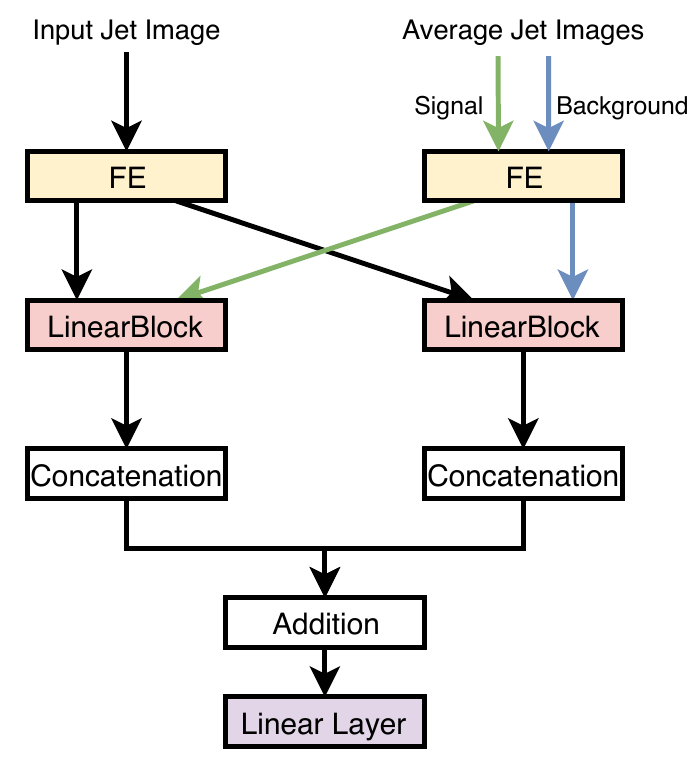}
        \label{CNN-1}}
    \subfigure[CNN-0]{
        \includegraphics[scale=0.8]{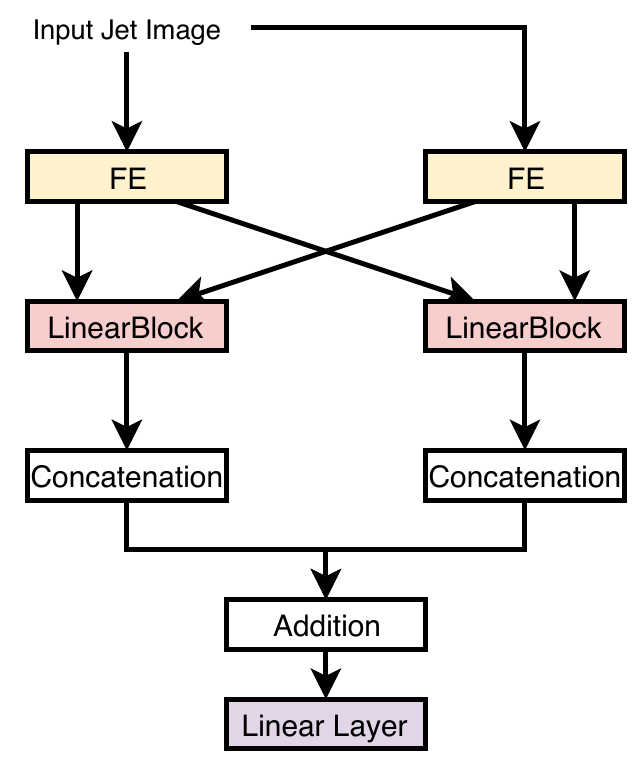}
        \label{CNN-0}}
    \caption{Three different architectures. Based on three FEs, we denote all networks as: 
             ABNN1, 2, 3 for \subref{ABNNs}, 
             CNN11, 21, 31 for \subref{CNN-1},
             CNN10, 20, 30 for \subref{CNN-0}.}
    \label{fig:ABNNs CNN-1 CNN-0}
\end{figure}

The feature maps from ConvBlock$4$, $7$, $10$ are viewed as the outputs. For the input jet image, they're the key-value pairs in different dimensions. For the average jet images, different queries in corresponding dimensions are generated. The dimension here means the output size, i.e., $(16,16,16)$, $(32,8,8)$, $(64,4,4)$ respectively (they are shown in (channel number, width, height)). Based on FEs, we developed ABNNs. Fig.\ref{ABNNs} shows their structures.

Key-value pairs and corresponding queries are passed to the AttnBlock and two kinds of attention are obtained, i.e., one is from the signal and the other is from the background. Each kind includes three attentions in 3 dimensions as mentioned. Attentions from the same kind are then concatenated into one compounded attention. We add them together and pass it to the final linear layer. Unified attention is expected to generated through these operations. We also set up CNN-1 networks shown in Fig.\ref{CNN-1}, which exclude the attention mechanism. This structure contains LinearBlocks instead of AttnBlocks, which is for comparison and showing the power of the attention. Upon CNN-1, CNN-0 structure in Fig.\ref{CNN-0} further removes the average jet images as additional inputs. This structure is a CNN made of two pipelines. Since 3 FEs can be chosen,  we denote the related networks as: ABNN1, 2, 3, CNN11, 21, 31, and CNN10, 20, 30.

\begin{table}[htp]
\begin{tabular}{|c|c|c|c|c|}
\hline
Scheme & Learning Rate & Gamma & Step Size & Patience \\ \hline
$1$    & $0.0001$      & $0.8$ & $20$      & $20$     \\
$2$    & $0.0001$      & $0.5$ & $20$      & $20$     \\
$3$    & $0.0001$      & $0.1$ & $20$      & $20$     \\
$4$    & $0.0001$      & $0.1$ & $10$      & $10$     \\ \hline
\end{tabular}
\caption{Training schemes.}
\label{table:training schemes}
\end{table}

All the networks mentioned in this article are written by PyTorch\cite{paszke2017}, and we also use Skorch\cite{skorch}, a scikit-learn\cite{scikit-learn} style PyTorch package to make the parameter tuning process easier. In the training process, in order to suppress over-fitting, we add a dropout layer after the ConvBlock40 layer with a $50\%$ dropout rate. We use the learning rate that decreases with the training process instead of a fixed one. It's well know that, if it is too large, the network tends to converge to the sub-optimal point; if it is too small, training may last too long (Early stopping is also used to prevent this happens). We use StepLR, a learning scheduler that multiplies the initial learning rate by a coefficient gamma every certain step size (epoch) to achieve the purpose of decreasing. By combining all the parameters: the patience of early stopping, gamma and step size parameters in stepLR, 4 training schemes in Table \ref{table:training schemes} are used.

\section{SAMPLE GENERATION}
\label{III}

We select the process of hardronically boosted Z decaying to dijet as signals and general QCD jets as backgrounds. Events are generated by Pythia8\cite{Sjoestrand2014} in pp collision at $\sqrt{s}=13$ TeV. Only the processes containing light quarks are chosen, ignoring the potential performance improvement brought by heavy quarks. We set the minimum $p_T$ cut $400$ TeV to generate highly collinear events in the standard model. All final state particles are kept on the condition that their pseudorapidity $|\eta| < \pi/2$. The decay products of $Z$ bosons as well as backgrounds are then clustered into jets by FastJet\cite{Cacciari2011} with the anti-$k_T$ algorithm\cite{Cacciari2008} using a distance parameter $R=1.0$. Subjets are generated by clustering all constituents of the corresponding jet using $k_T$ algorithm\cite{Catani1993, Ellis1993} with $R= 0.3$. We only keep subjets whose $p_{T}^{\text {subjet }}<0.05 \times p_{T}^{\text {jet }}$ to mitigate the contribution from the underlying events. We set the resolution $\Delta \eta=\Delta \phi=0.1$ and generate jet images of the size $32 \times 32$, which follow the configuration in \cite{Almeida2015} and \cite{Oliveira2015}. We take a similar preprocessing procedure in \cite{Oliveira2015}: translation, pixelation, rotation. No flip and normalization are performed since we do not observe any benefit of them to improve the performance. Note the normalization here means scaling the range of pixel intensities into $[0,1]$ with the max intensity. Average jet images, the additional inputs, are generated by taking the sum of all jet images in the training set and performing normalization. We show them in Fig.\ref{fig:Average jet images}.

The dataset consists of 500,000 jets in total. 300,000 jets are used to training the networks and 100,000 for validation. The test set contains the same amount of jets as validation.

\section{RESULT ANALYSIS}
\label{IV}

\begin{figure}[htp]
    \centering
    \subfigure[]{
        \includegraphics[scale=0.5]{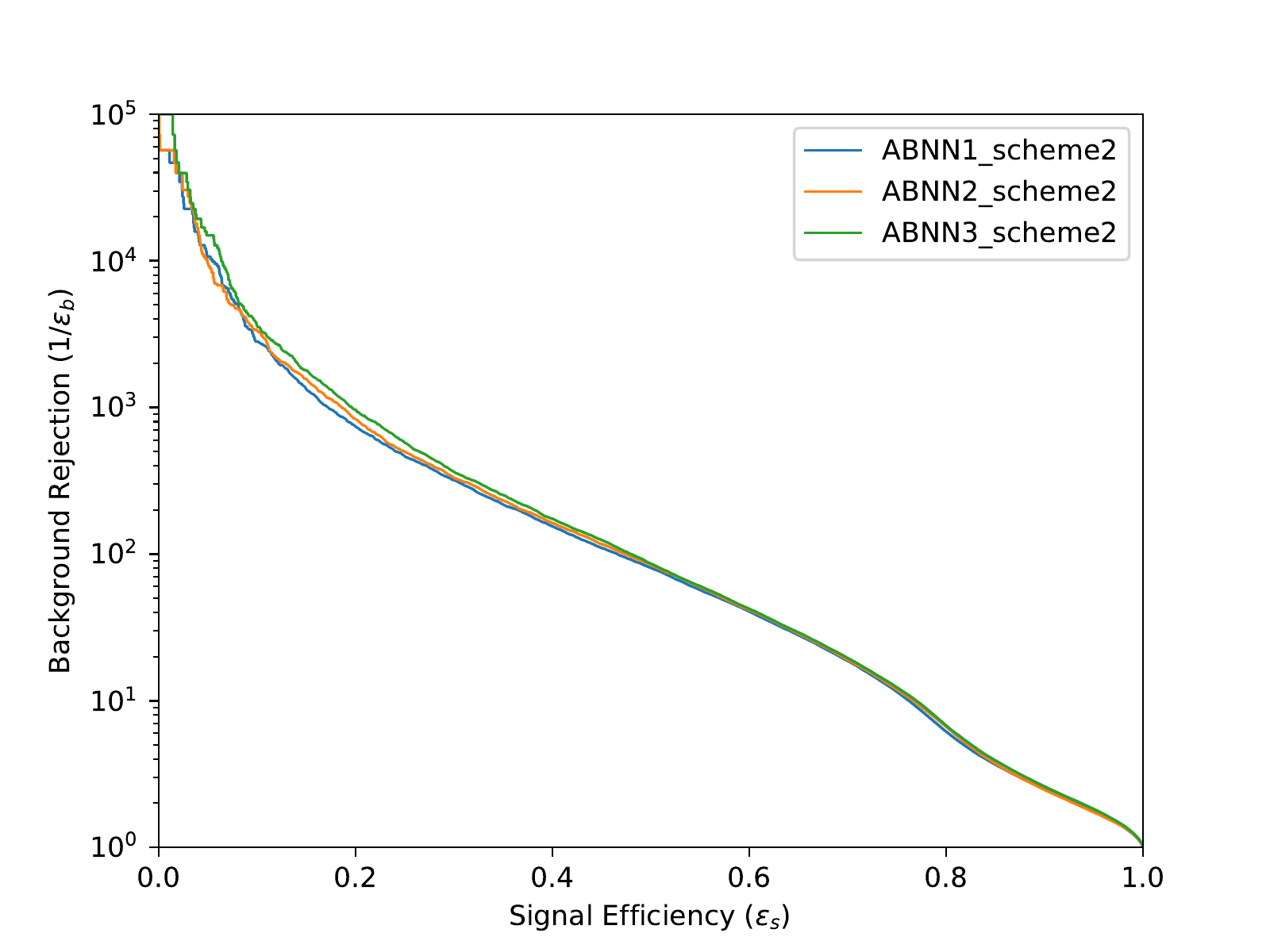}
        \label{subfig: roc ABNNs}}
    \subfigure[]{
        \includegraphics[scale=0.5]{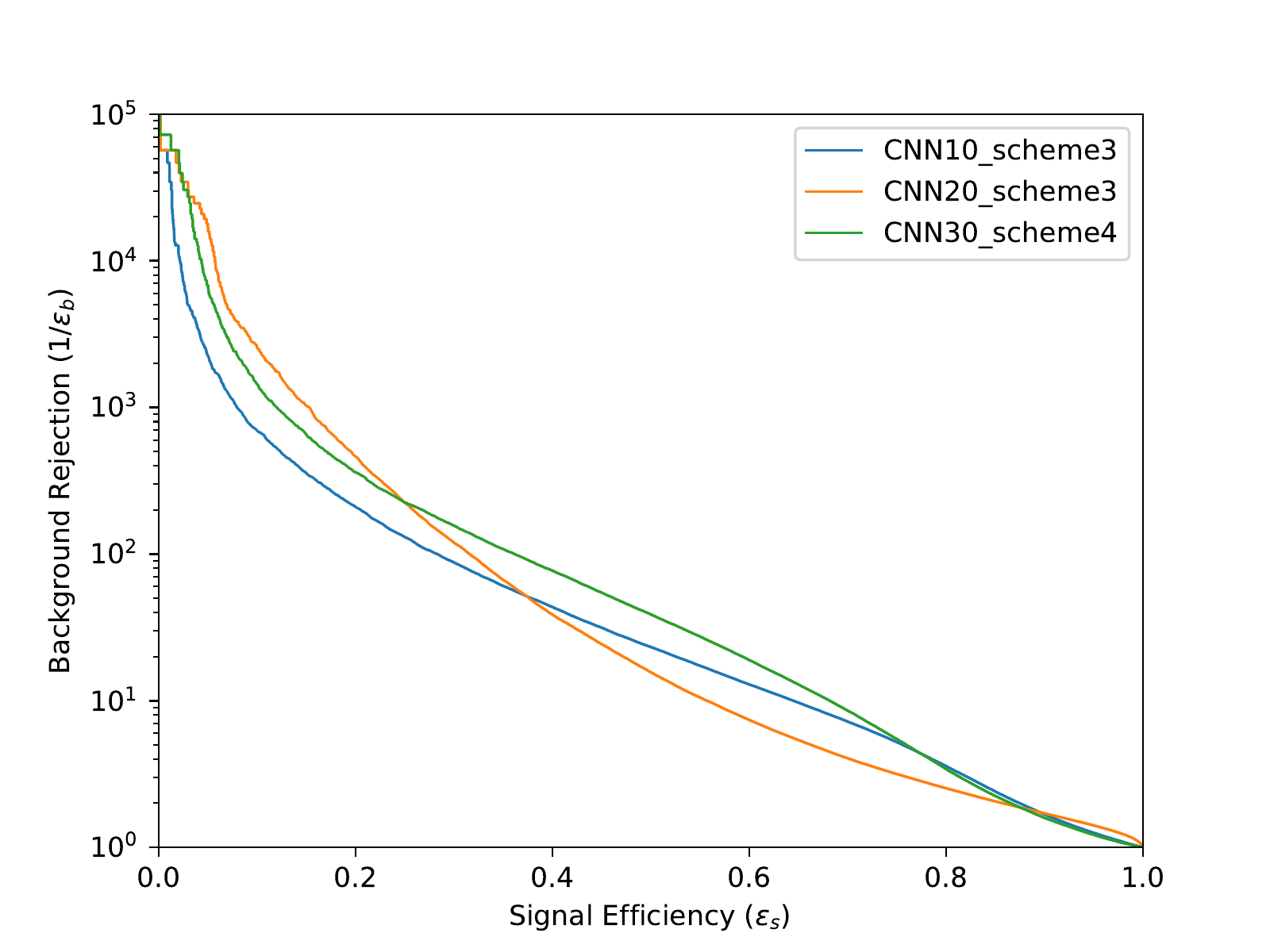}
        \label{subfig: roc CNN-0}}
    \subfigure[]{
        \includegraphics[scale=0.5]{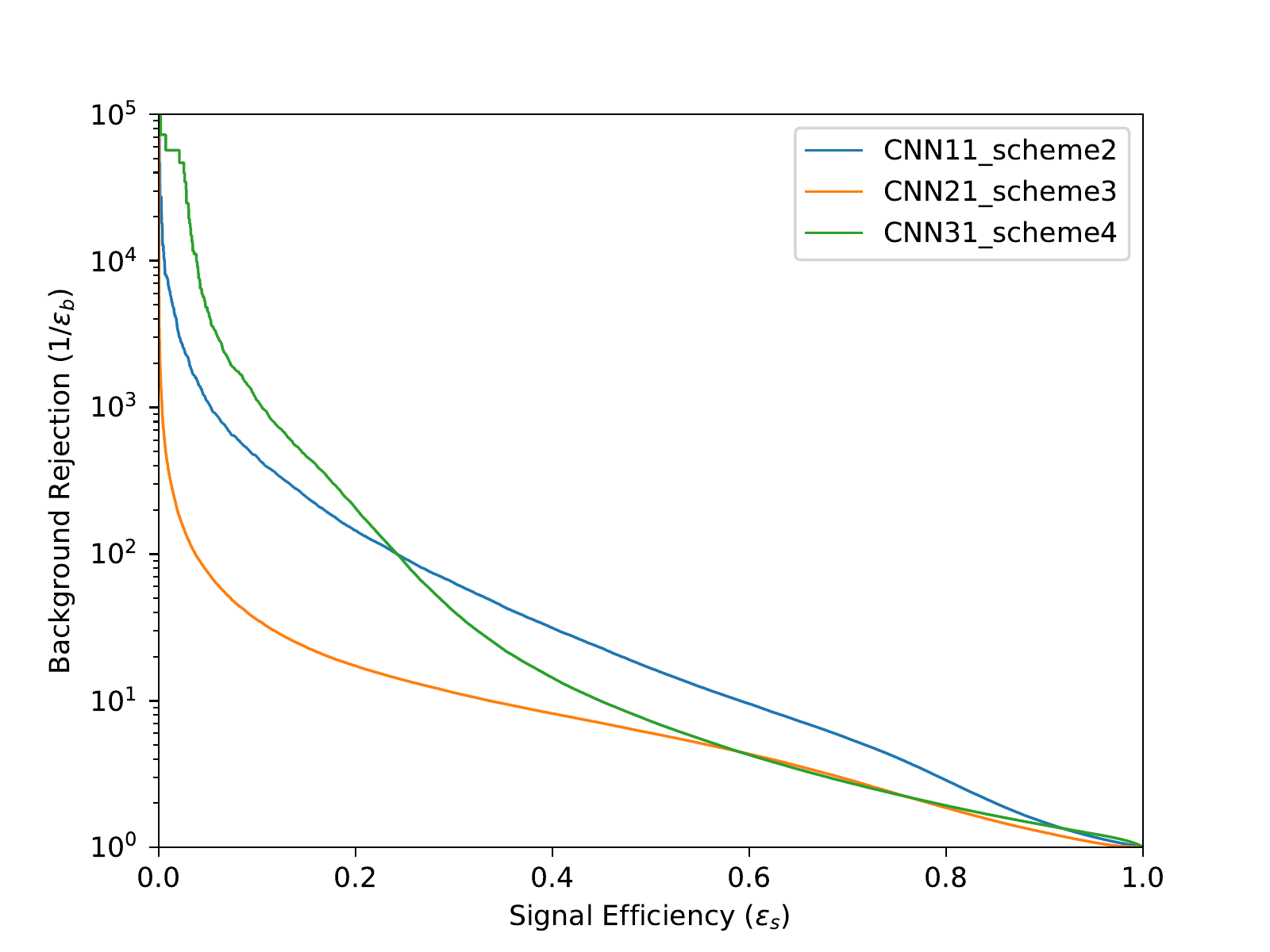}
        \label{subfig: roc CNN-1}}
    \caption{ROC curves of three architectures:
             \subref{subfig: roc ABNNs}: ABNNs, 
             \subref{subfig: roc CNN-0}: CNN-0, 
             \subref{subfig: roc CNN-1}: CNN-1.
             Only the networks in most efficient schemes are shown.}
    \label{fig:ROC curves of three kinds of architectures}
\end{figure}

The receiver operating characteristic (ROC) is used to illustrate the classification ability as the signal threshold is varied. We denote the signal efficiency as $\epsilon_s$ and the mis-tag rate as $\epsilon_b$. The reciprocal mis-tag rate $1/\epsilon_b$, also known as the background rejection, is plotted on the vertical axis. Consequently, the higher the curve, the better the performance. Other than the curve, three typical metrics for evaluation are used: the area under the ROC curve (AUC), the best accuracy in the training process (ACC), and the background rejection at a signal efficiency $50\%$ ($1 /\epsilon_{b}|_{\epsilon_{s}=50 \%}$, denoted by R50).

\begin{table}[htp]
\begin{tabular}{lccc}
\hline
      & R50     & AUC    & ACC      \\ \hline
CNN30 & 16.6189 & 0.8094 & $0.8313$ \\ \hline
CNN11 & 38.7689 & 0.8386 & $0.8232$ \\ \hline
ABNN1 & 80.4809 & 0.8965 & $0.8300$ \\ \hline
\end{tabular}
\caption{Performance metrics for the best networks.}
\label{table:Performance metrics for the best networks}
\end{table}

\begin{figure}[htp]
    \centering
    \includegraphics[scale=0.5]{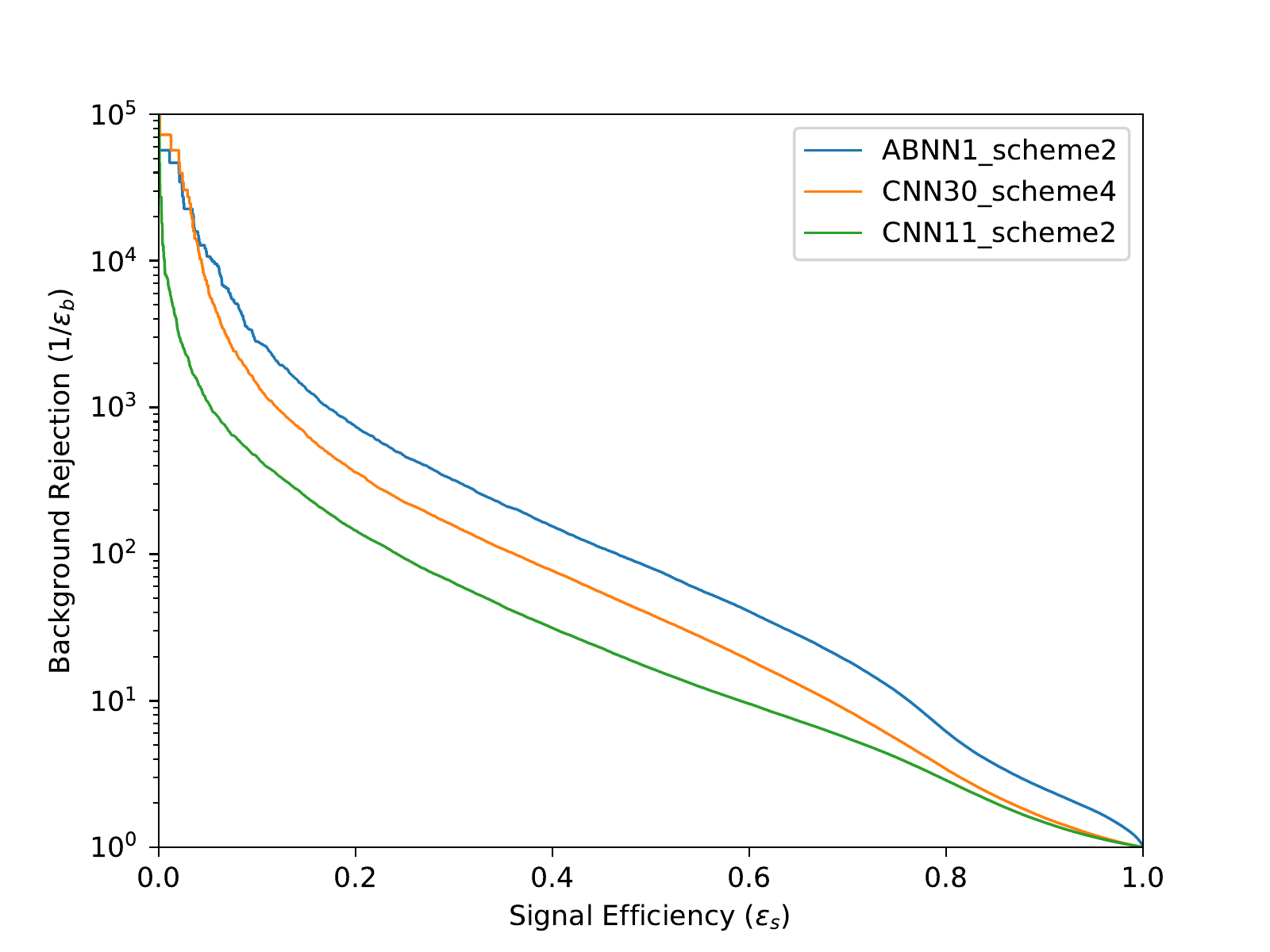}
    \caption{Performance comparison of the best networks}
    \label{fig:ROC Performance comparison of the best networks}
\end{figure}

Through training under different schemes, in Fig.\ref{fig:ROC curves of three kinds of architectures} we have made ROC curves of three network architectures trained in the most efficient schemes. It can be seen that the deeper networks does not always perform better. We presume that under this design, the deepened network becomes too complicated for the given data, leading to the decline of generalization power. Among the three kinds of architectures, we select the best networks to compare their performance. The ROC curves are shown in Fig.\ref{fig:ROC Performance comparison of the best networks} and performance metrics are in the Table \ref{table:Performance metrics for the best networks}. 

\subsection{Comparison of CNN-0, CNN-1, and ABNNs}
\label{IV.A}

By comparing CNN11 and CNN30, in the absence of the attention mechanism, the performance of CNN11 is reduced due to the influence of the additional input average jet images. The reason for this is that we do not input the corresponding average jet image according to its category, but input both categories of average jet images. This operation is equivalent to adding an ordinary feature that is the same between signal and background jet images, which cannot improve the classification ability of the network. After introducing the attention mechanism, ABNN1 has achieved better results than CNN30, and the improvement in background rejection is about R50. This shows obviously that for an input jet image, the network successfully notices its corresponding relationship with two average jet images. That is which part of the input jet image should be paid attention to. Such attention effectively improves the discrimination ability of the network. The attention maps are shown in the Chap.\ref{V}, which are more visually straightforward.

\subsection{Stacked CNNs}
\label{IV.B}

Our neural network passes the inputs through two independent feature extractors, and then combines the feature maps to output the final classification results. Although ABNN1 has achieved better performance than CNN30 and CNN11 by introducing the attention mechanism, we are also curious about its performance compared with standard stacked CNNs. In order to explore stacked CNNs with similar performance to our proposed structure, we designed four stacked CNNs as shown in the Table \ref{table:stacked CNNs}.

\begin{table}[htp]
\begin{tabular}{|c|c|c|c|c|}
\hline
            & CNN1                     & CNN2                     & CNN3                     & CNN4                      \\ \hline
ConvBlock1  & $1 \to 8$                & $1 \to 8$                & $1 \to 8$                & $1 \to 8$                 \\
Max Pooling &                          &                          &                          &                           \\
ConvBlock2  & $8 \to 16$               & $8 \to 16$               & $8 \to 16$               & $8 \to 16$                \\
Max Pooling &                          &                          &                          &                           \\
ConvBlock3  &                          & $16 \to 32$              & $16 \to 32$              & $16 \to 32$               \\
Max Pooling &                          &                          &                          &                           \\
ConvBlock4  &                          &                          & $32 \to 64$              & $32 \to 64$               \\
Max Pooling &                          &                          &                          &                           \\
Dropout     &                          &                          &                          &                           \\
ConvBlock5  &                          &                          &                          & $64 \to 128$              \\
Max Pooling &                          &                          &                          &                           \\
Dropout     &                          &                          &                          &                           \\ \hline
Linear      & $16\times8\times8 \to 2$ & $32\times4\times4 \to 2$ & $64\times2\times2 \to 2$ & $128\times1\times1 \to 2$ \\ \hline
\end{tabular}
\caption{Structures and channel variation of stacked CNNs.}
\label{table:stacked CNNs}
\end{table}

\begin{figure}[htp]
    \centering
    \includegraphics[scale=0.5]{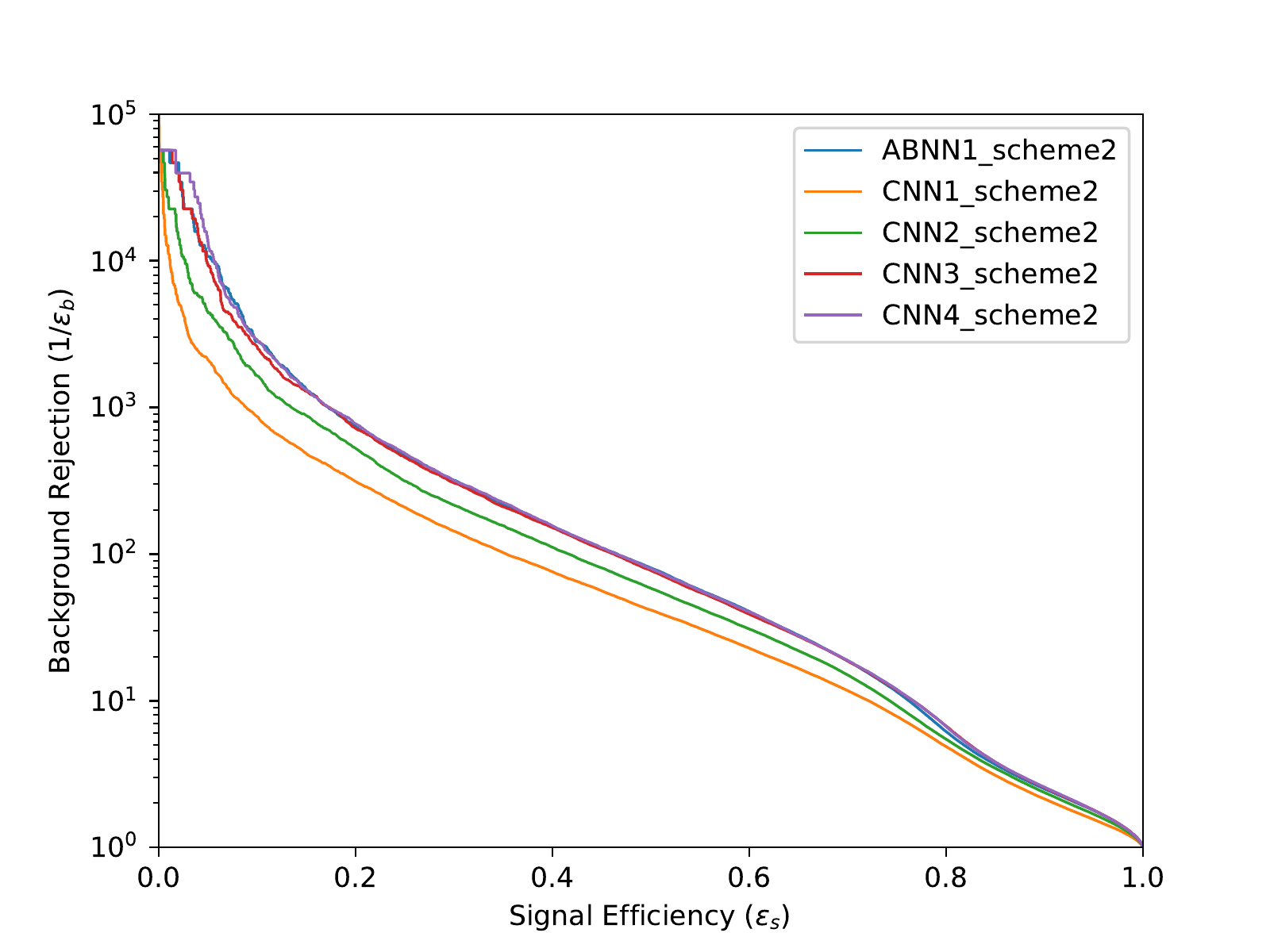}
    \caption{ROC curves of stacked CNNs.}
    \label{fig:ROC curves of stacked CNNs}
\end{figure}

The basic design idea is to start with the number of channels 16, and then add a ConvBlock in turn with the number of channels as twice as the previous one. One max pooling layer follows every ConvBlock. Up to 5 ConvBlocks, the resulting feature map has only one pixel. In order to keep the experiment simple and not change the basic block, we did not explore the possibility of a deeper network, although \cite{Chen2019} has designed CNNs with a maximum number of channels of 256. To obtain a deeper network, you can change the corresponding kernel or padding number of the convolutional layer in the ConvBlock. This limits the decrease of the feature map size, so you can stack more layers. We train these four networks with the same strategy as mentioned in Chap.\ref{II}. The Fig.\ref{fig:ROC curves of stacked CNNs} is the ROC curve obtained.

\begin{figure}[H]
    \centering
    \includegraphics[scale=0.5]{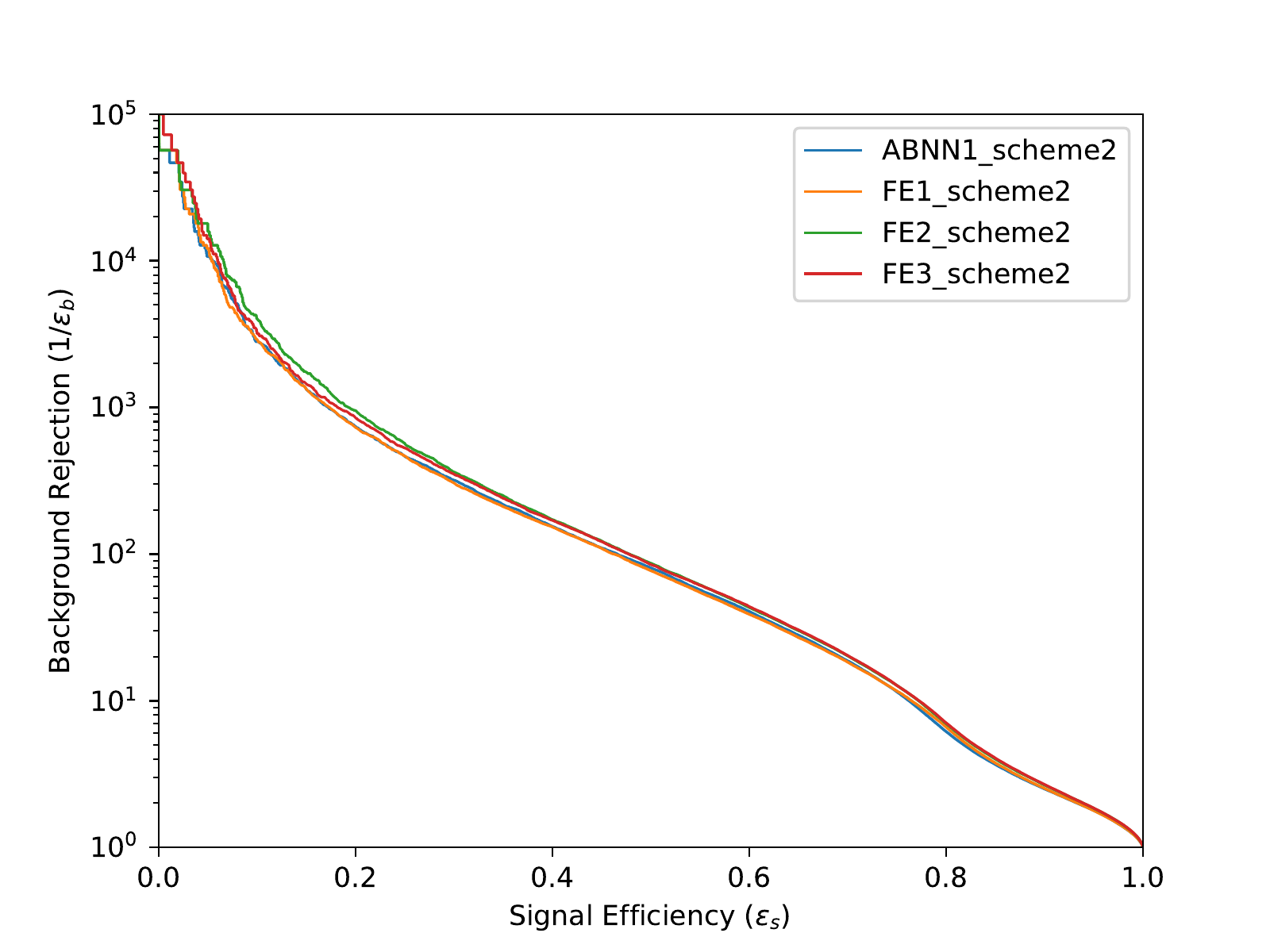}
    \caption{ROC curves of feature extractors}
    \label{fig:ROC curves of feature extractors}
\end{figure}

The deeper CNN4 does not get better results as expected, and its performance is almost as good as CNN3 and ABNN1. Although in general, deeper networks have stronger resolving power, which is only for data with potentially sufficiently complex structures. This shows that for our data, networks such as CNN3 have reached the upper limit of their recognition capabilities. If we only continue to increase the network depth, it will not bring additional recognition capabilities. We noticed that the difference between CNN3 and FE1 only exists in the output, so by changing the output, we also use the above FE2, 3 as stacked CNNs to explore the difference between their recognition capabilities and that of the ABNN1. It is worth noting that FEs increase the depth by increasing the number of ConvBlock whose channel number is unchanged. Not every ConvBlock is followed by a max pooling layer, which is different from the stacked CNNs designing idea. The Fig.\ref{fig:ROC curves of feature extractors} shows the ROC curves obtained.

All the curves in the figure basically overlap, which means that their performance is basically the same. This also indicates that CNN3 is the simplest stacked CNN for our data. Whether increasing the number of ConvBlock with larger channel numbers or the same channel numbers, its performance has reached the upper limit. By comparing the ABNN1 and the CNN3, under our design, the ABNN1 has the number of parameters as nearly twice as the CNN3. Because they are very shallow compared with many networks in machine learning fields, the difference is not very big.

\subsection{Different Average Jet Images}
\label{IV.C}

\begin{figure}[htp]
    \centering
    \includegraphics[scale=0.35]{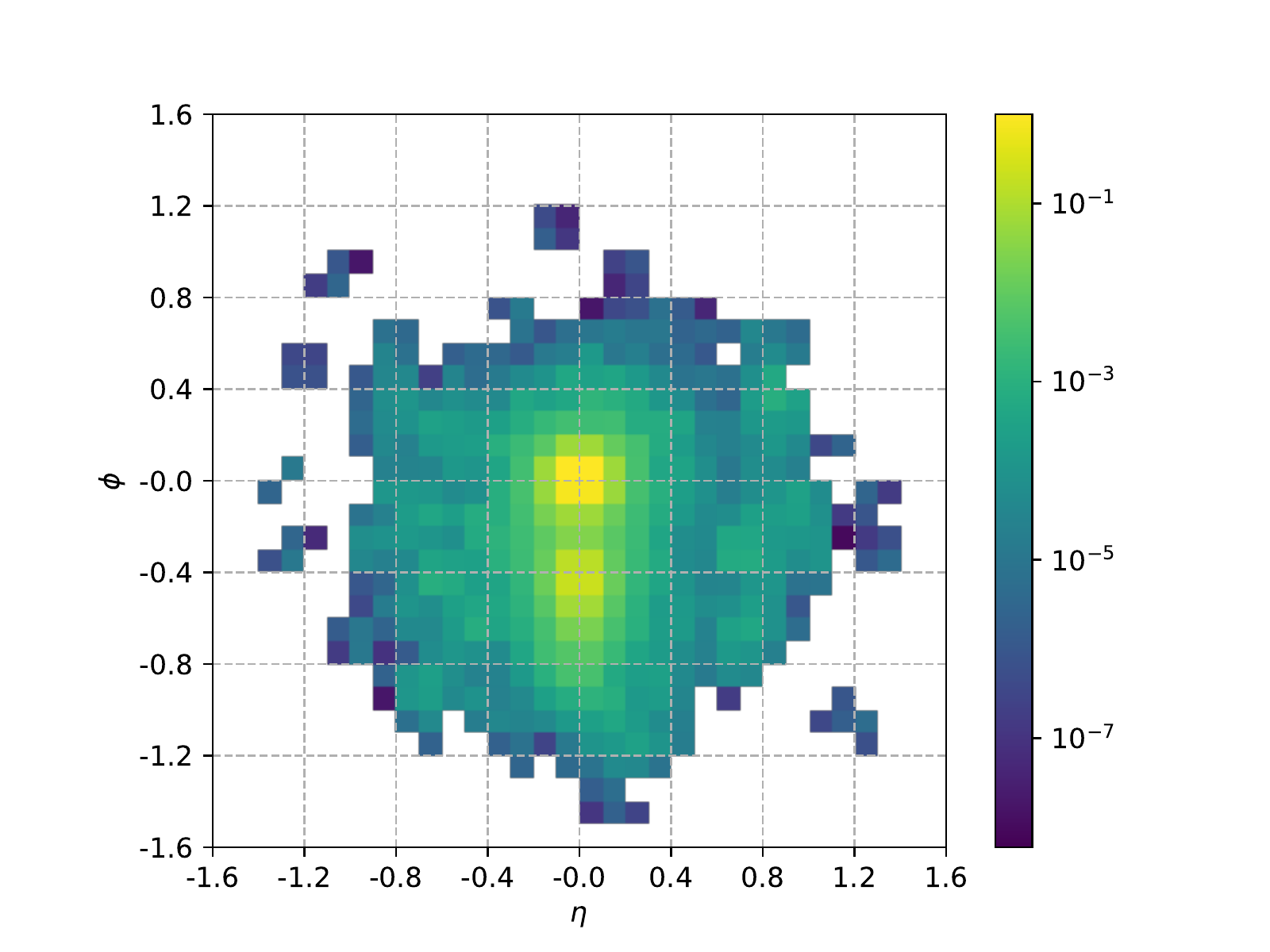}
    \includegraphics[scale=0.35]{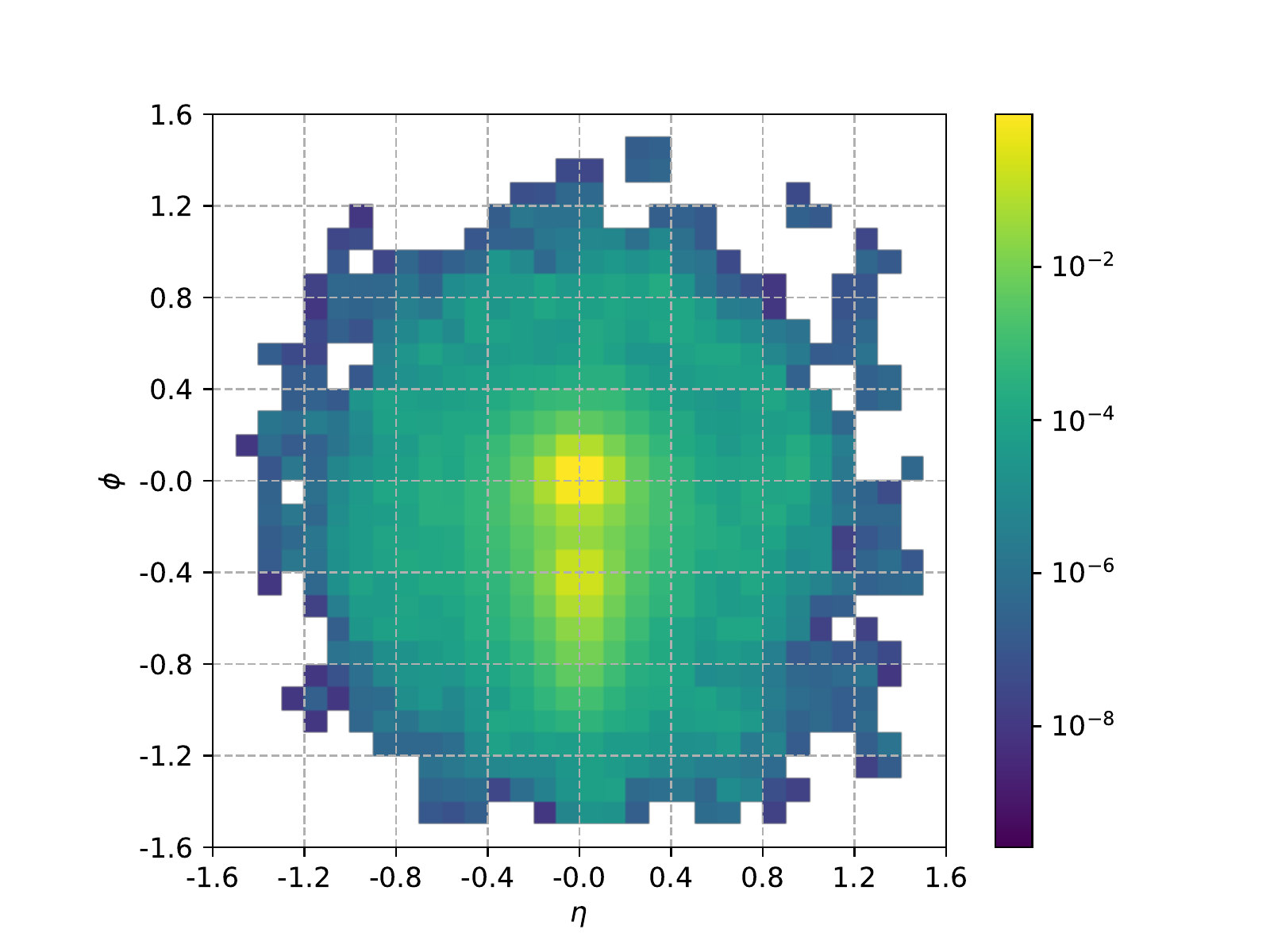}    
    \includegraphics[scale=0.35]{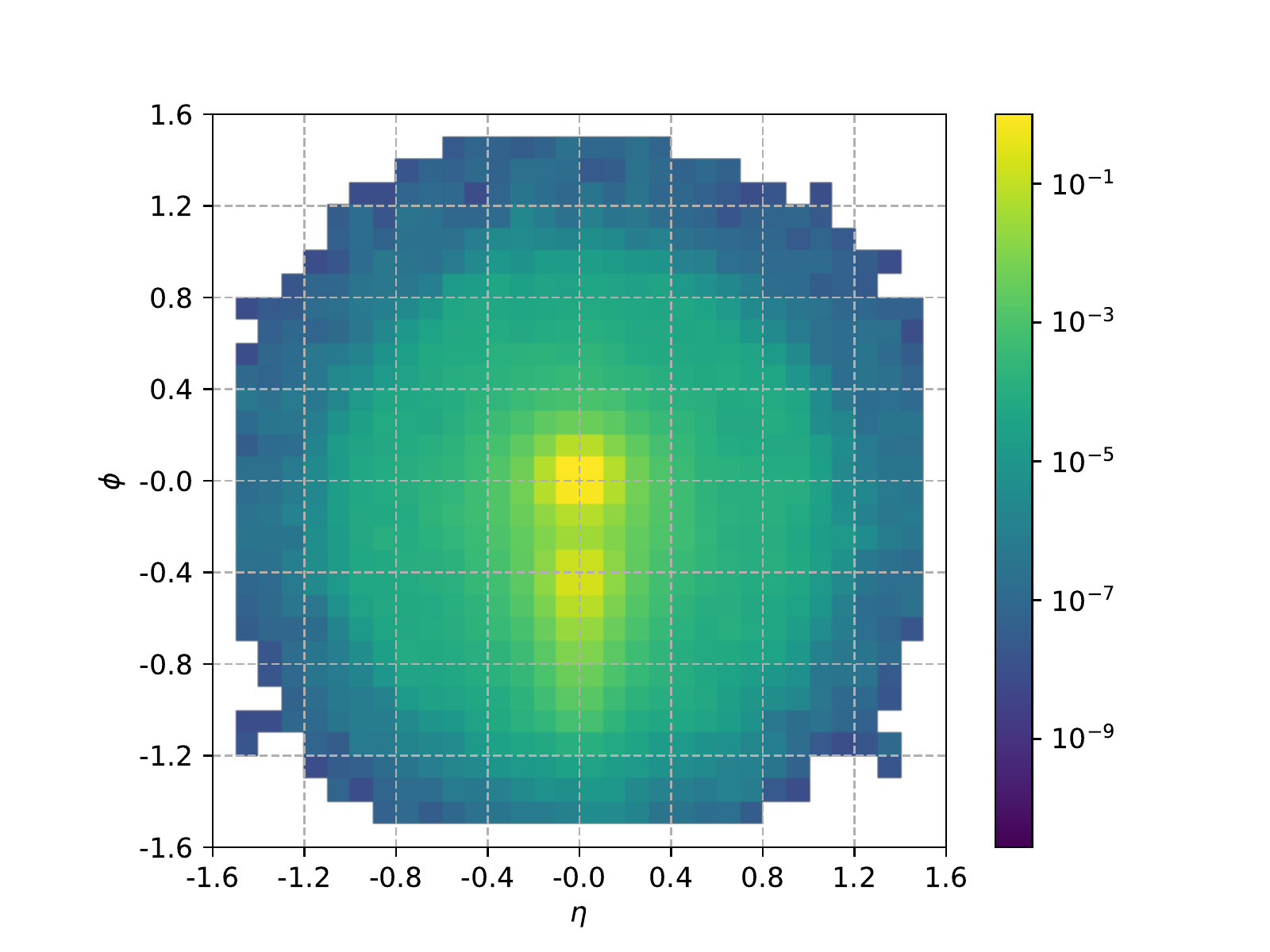}    
    \includegraphics[scale=0.35]{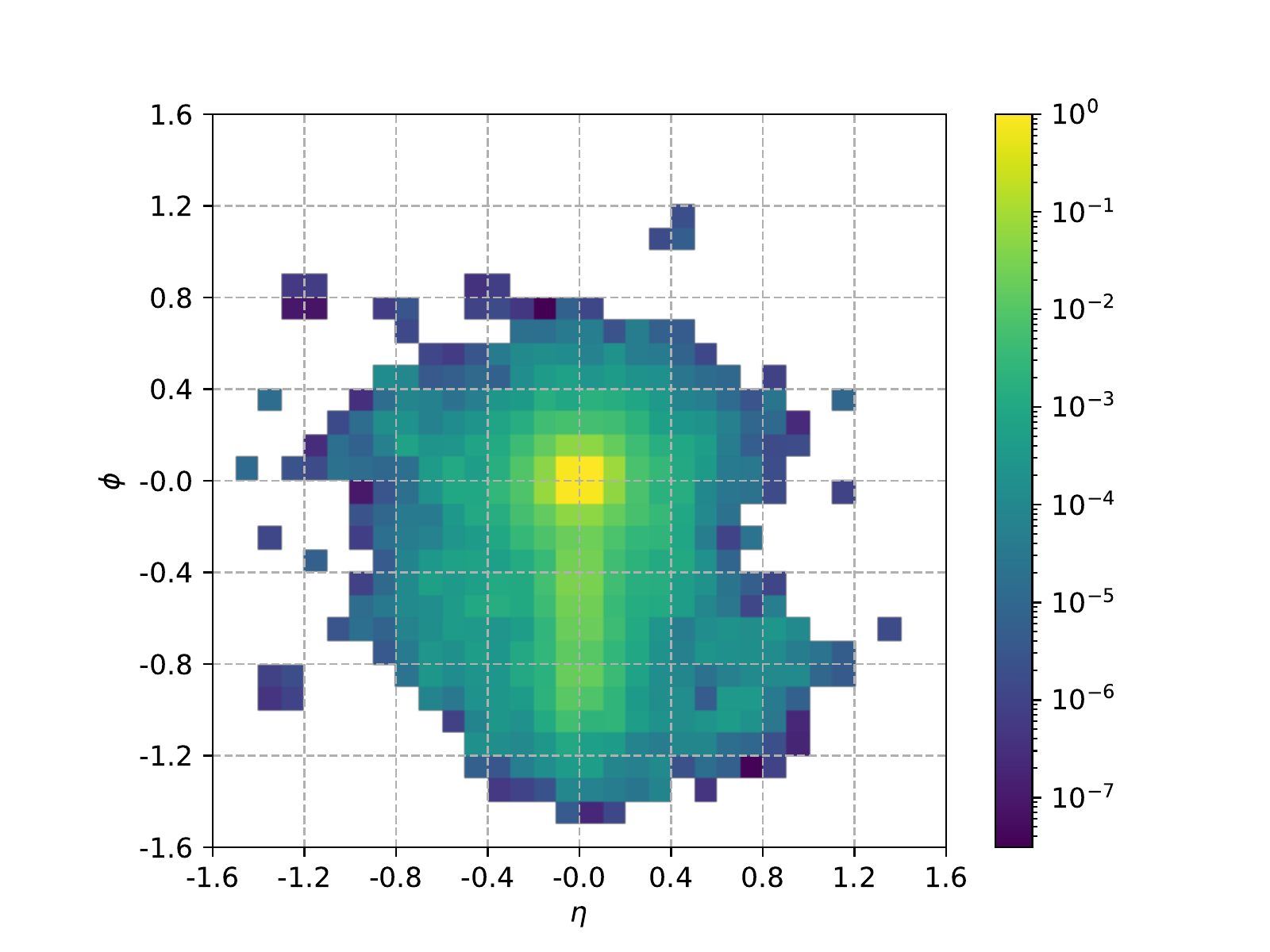}
    \includegraphics[scale=0.35]{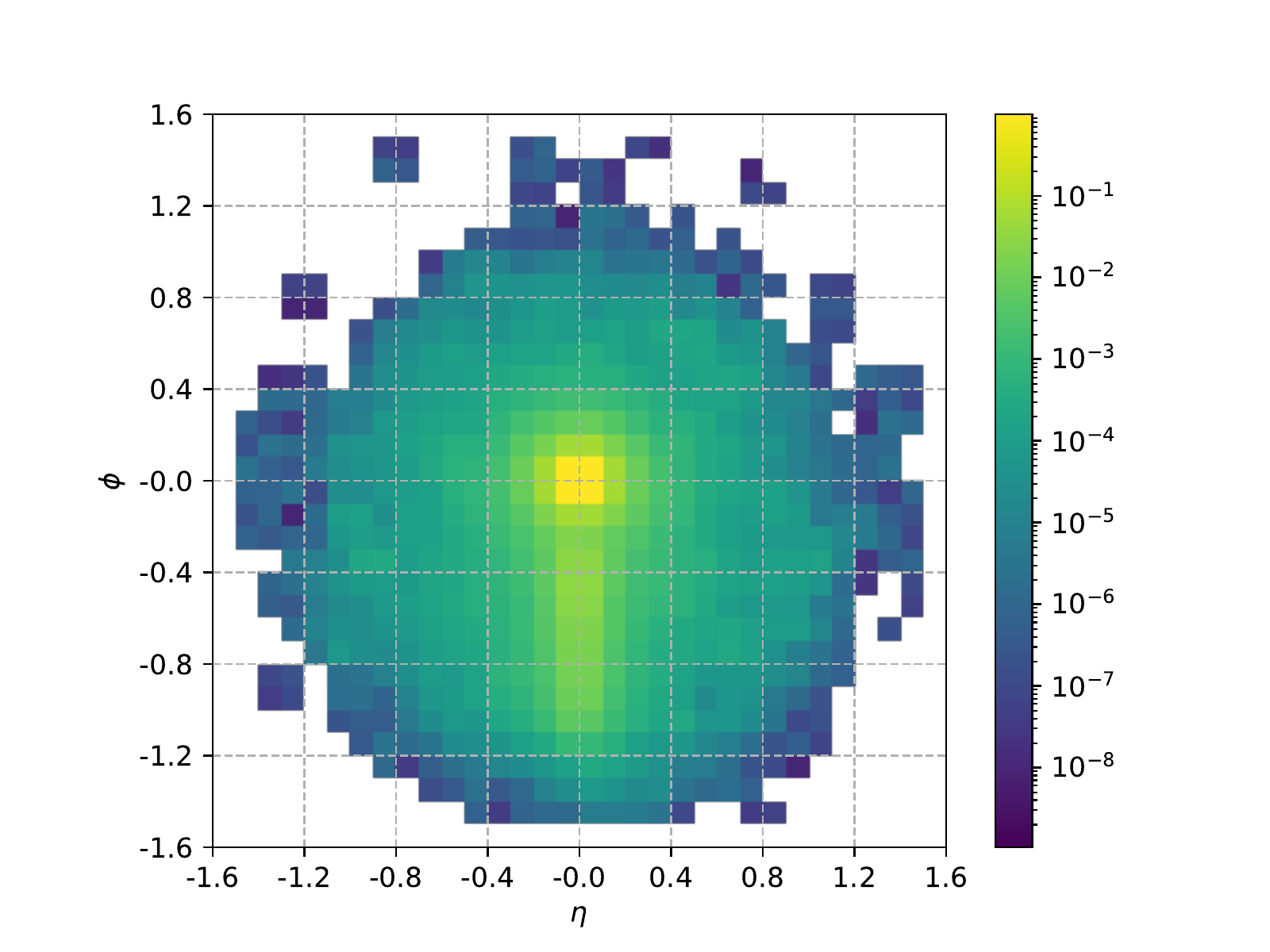}
    \includegraphics[scale=0.35]{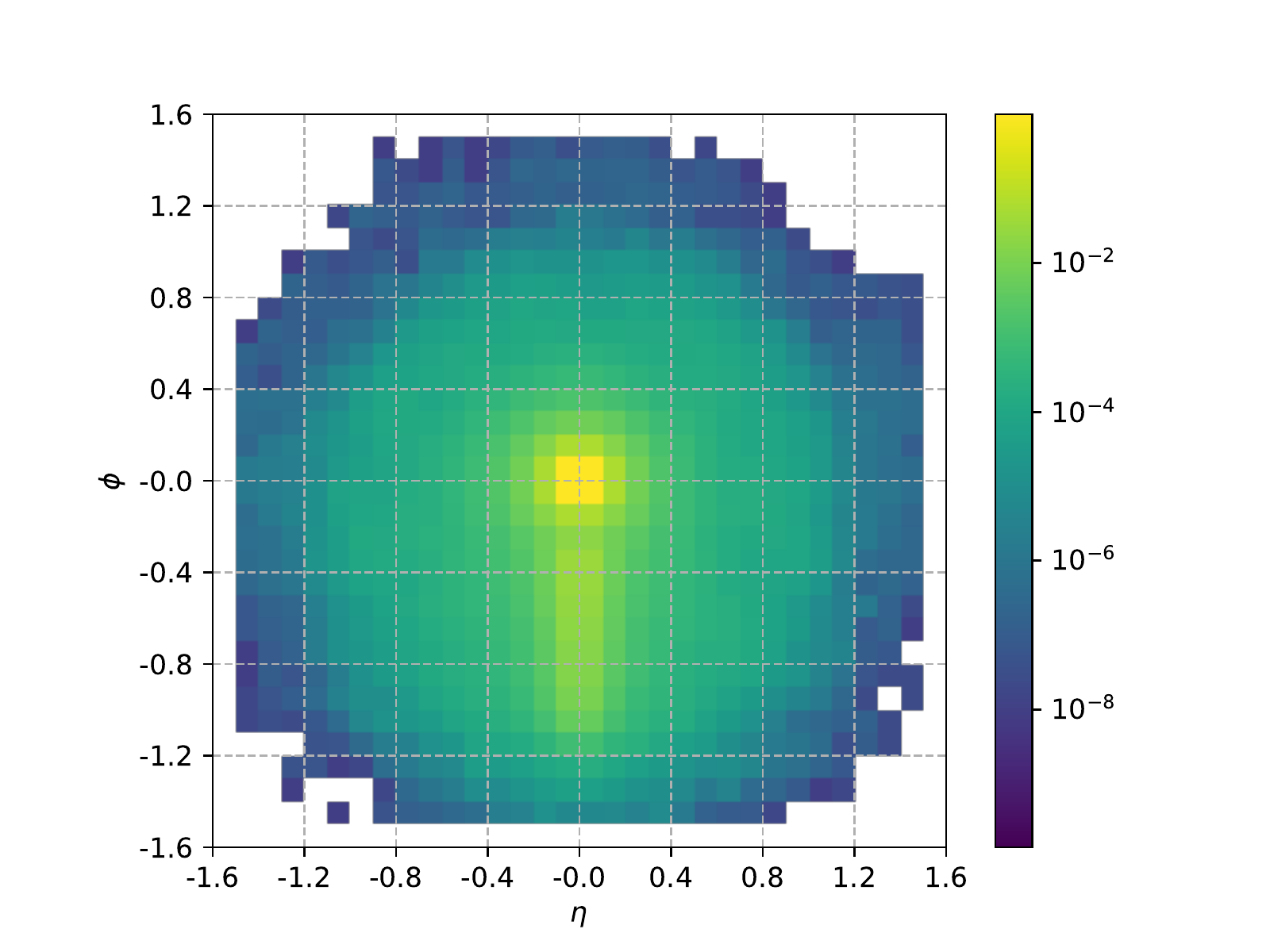}
    \includegraphics[scale=0.35]{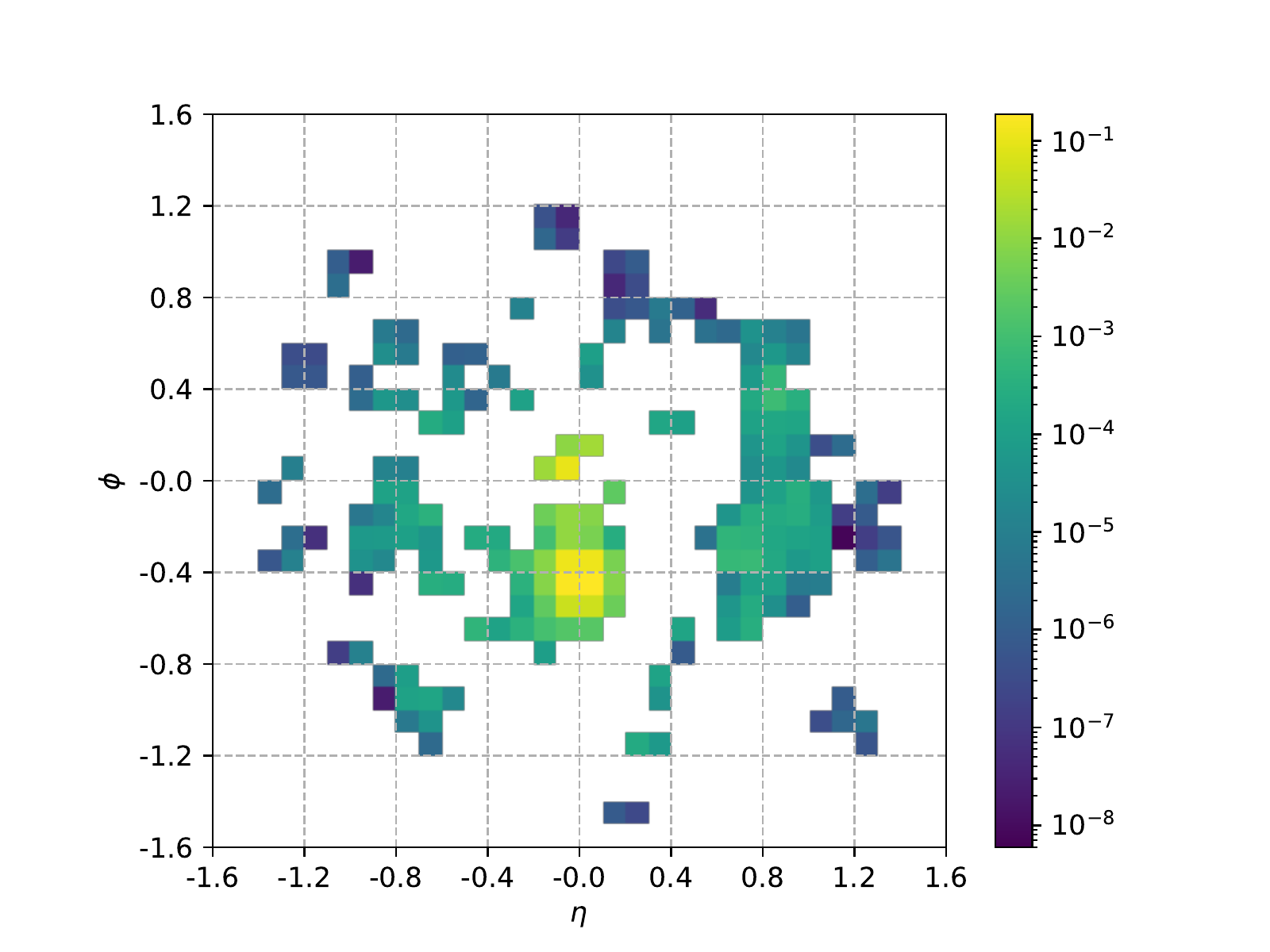}
    \includegraphics[scale=0.35]{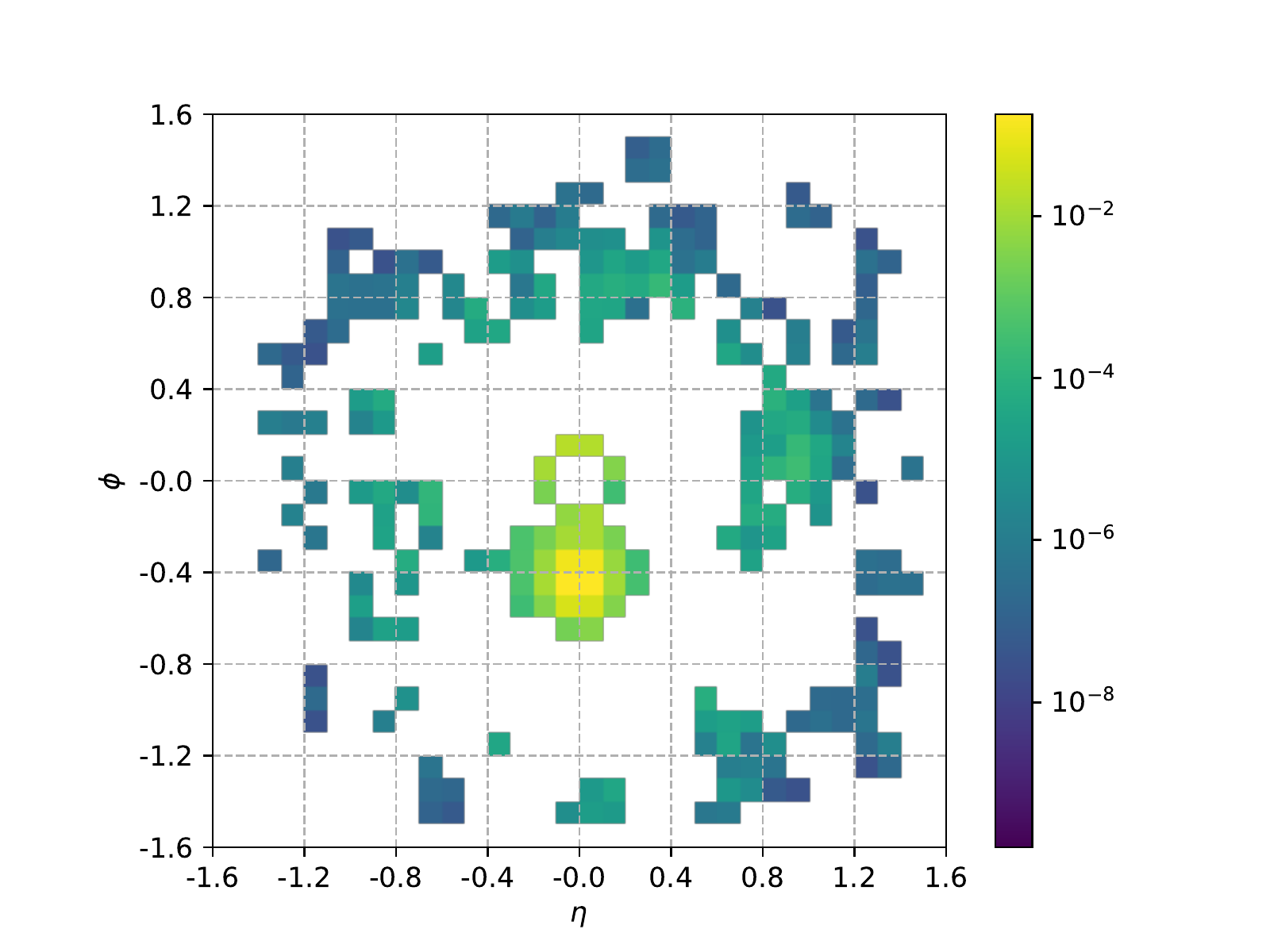}
    \includegraphics[scale=0.35]{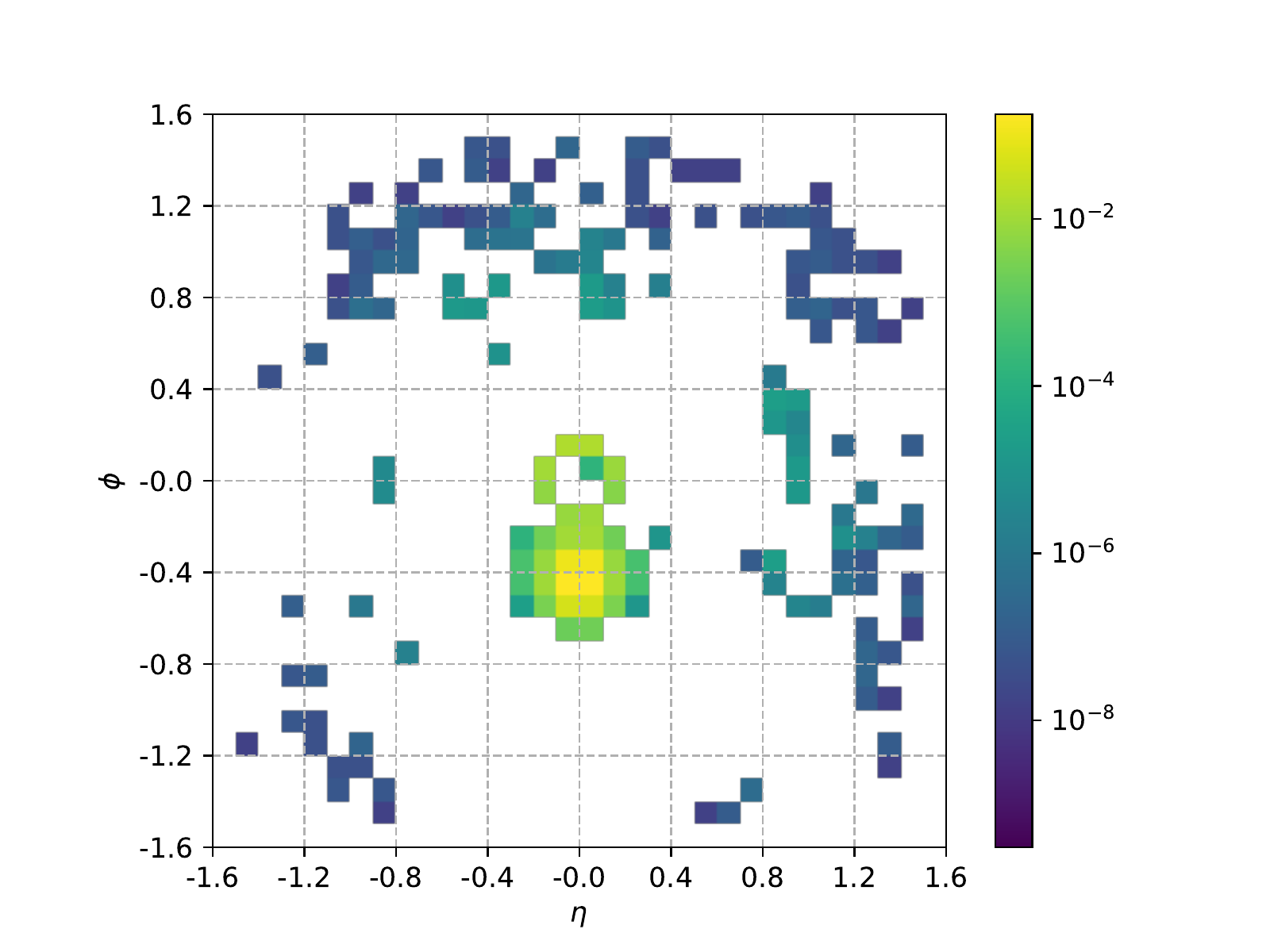}
    \caption{Different Average Jet Images.
             The columns represent the average jet images made of $1/1000, 1/100, 1/10$ jet images respectively.
             The rows represent the signal and background average jet images, and their difference respectively.}
    \label{fig:different average jet images}
\end{figure}

In this section we explore the impact of different average jet images on ABNNs. For ABNNs, we use the average jet images from the signal and the background as additional inputs. By using them as queries, the network learns the points needed to be paid attention to in the input jet image. In all previous experiments, the average jet images used by default consist of all jet images in the training set, as shown in Fig.\ref{fig:Average jet images}. As mentioned above, we believe that such an average jet image contains all the potential features, so it is the default choice. We also check the difference in average jet images composed of different numbers of jet images, as shown in Fig.\ref{fig:different average jet images}.

\begin{figure}[htp]
    \centering
    \includegraphics[scale=0.5]{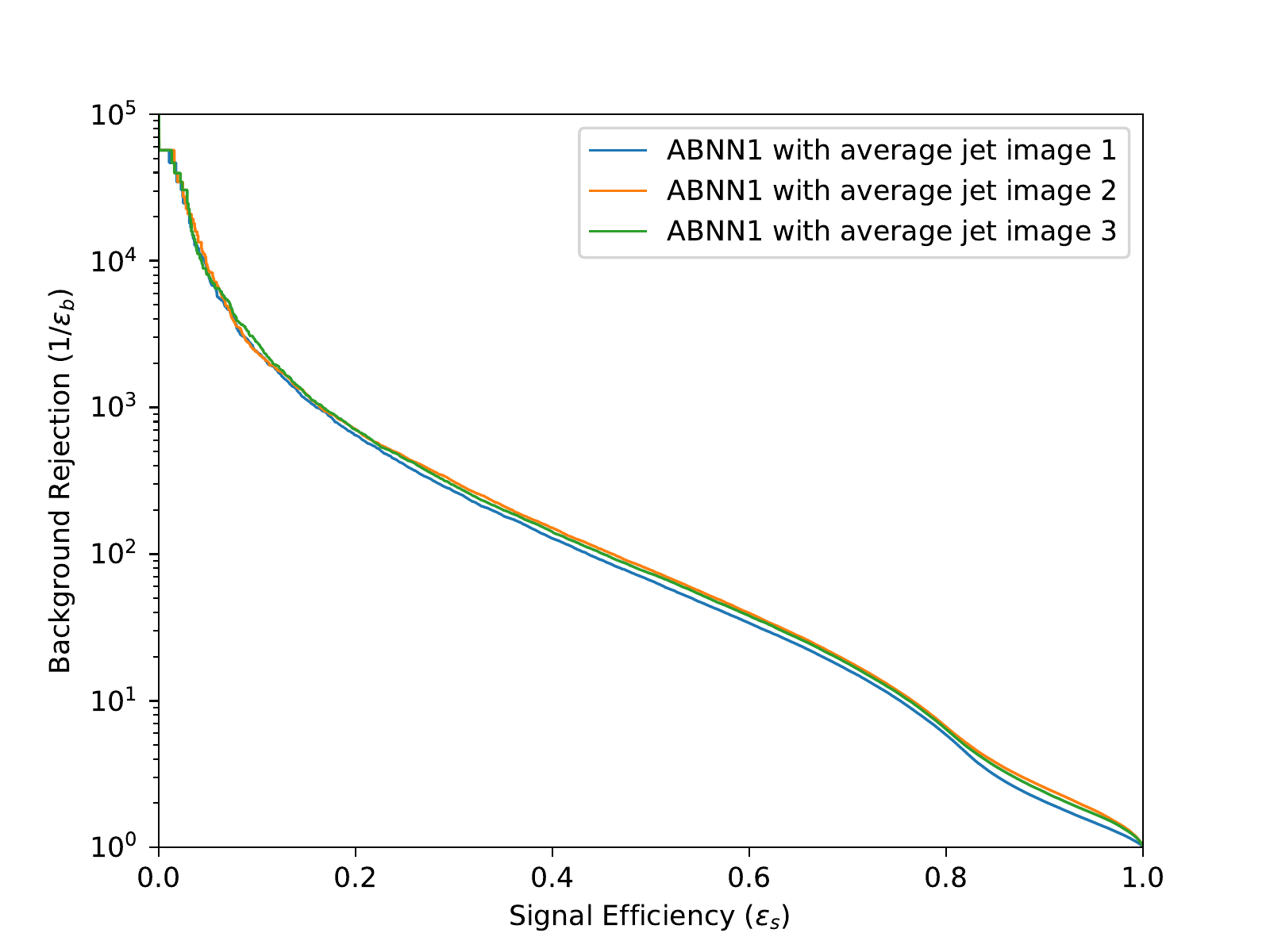}
    \caption{ROC curves of different average jet images. We use average jet image1-3 to denote  3 sets of the signal and background average jet images as mentioned.}
    \label{fig:ROC curves of ABNN1 with different average jet images}
\end{figure}

\begin{figure}[htp]
    \centering
    \includegraphics[scale=0.5]{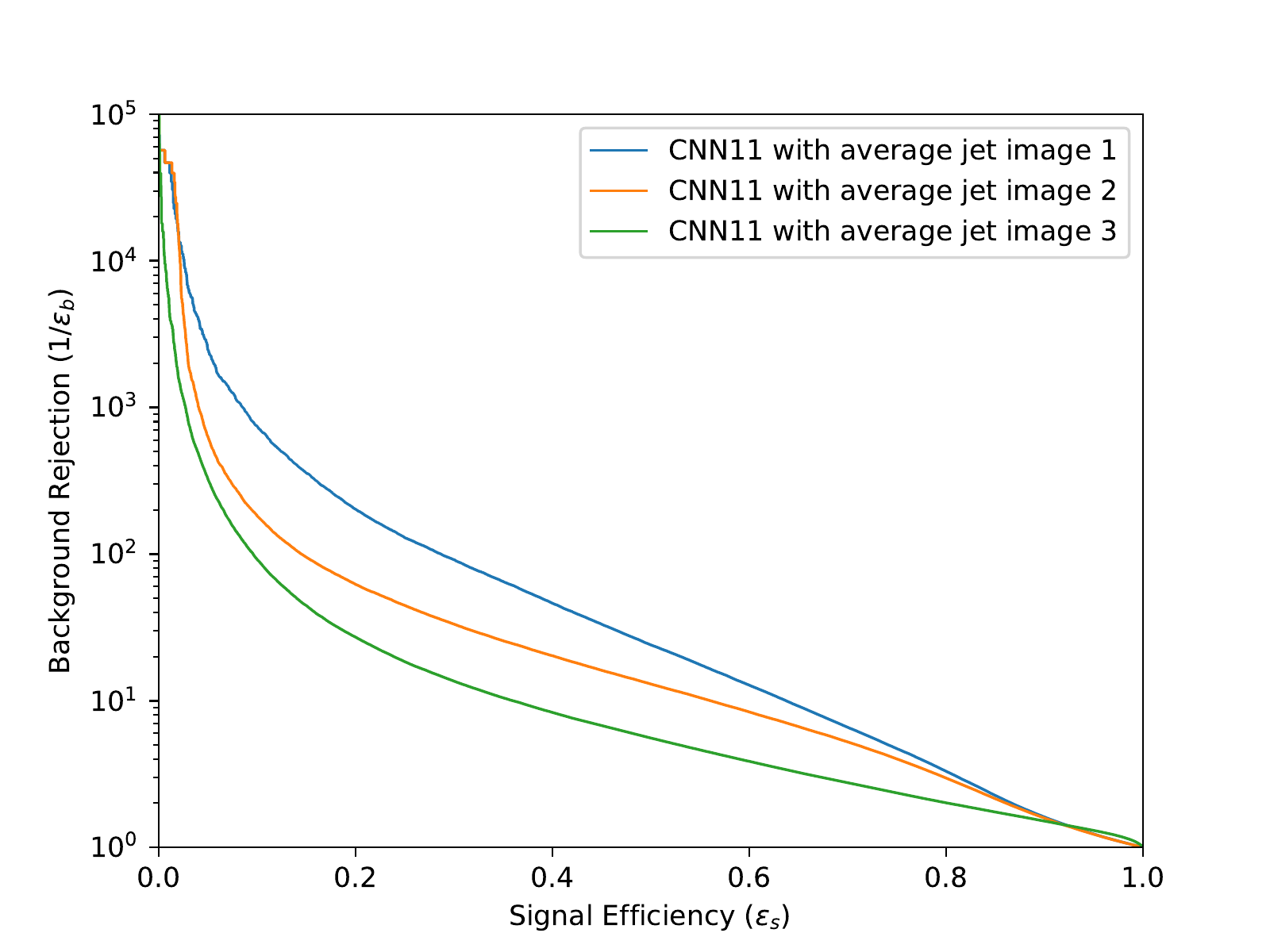}
    \caption{ROC curves of different average jet images. We use average jet image1-3 to denote  3 sets of the signal and background average jet images as mentioned.}
    \label{fig:ROC curves of CNN11 with different average jet images}
\end{figure}

The first two rows clearly show the changes in the average jet images as the number of jet images increases. The difference mainly lies in the surrounding area, while the pixel intensites at the location of the subjet is basically unchanged. We pass these different average jet images into ABNN1 which achieved the best performance before to observe their influence. CNN11 is also used to compare the consequence with ABNN1.  Their ROC curves are shown in Fig.\ref{fig:ROC curves of ABNN1 with different average jet images} and Fig.\ref{fig:ROC curves of CNN11 with different average jet images}. 

The ROC curves of ABNN1 also confirms the point: the differences in these surrounding areas do not affect the performance of our network. Unsurprisingly, for the CNN11 which does not have the attention mechanism, it takes these areas into account, resulting in a decrease in the performance.

\section{VISUALIZATION}
\label{V}
\begin{figure}[htp]
    \centering
    \subfigure[]{
        \includegraphics[scale=0.5]{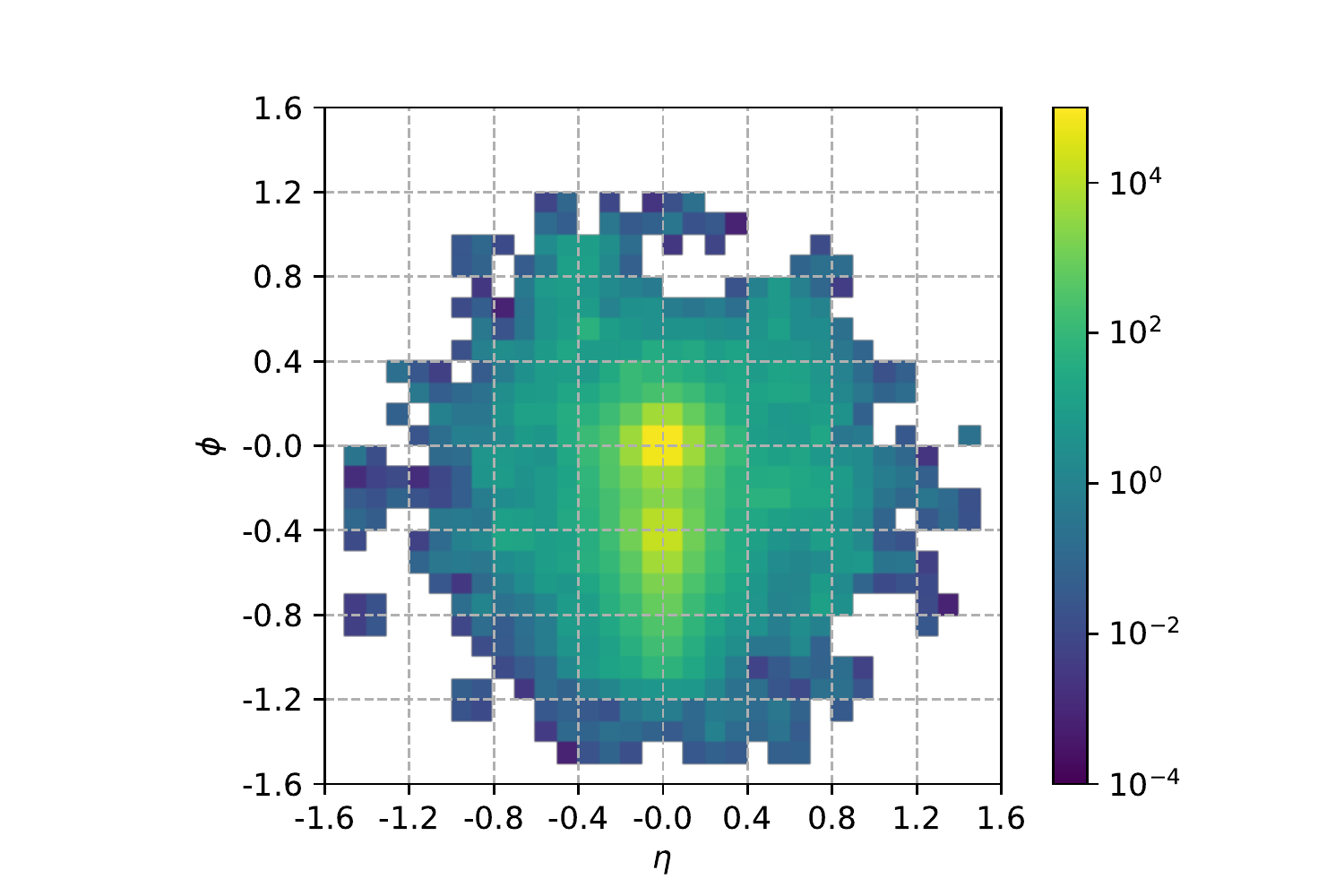}
        \label{1000 signal samples}
    }
    \subfigure[]{
        \includegraphics[scale=0.5]{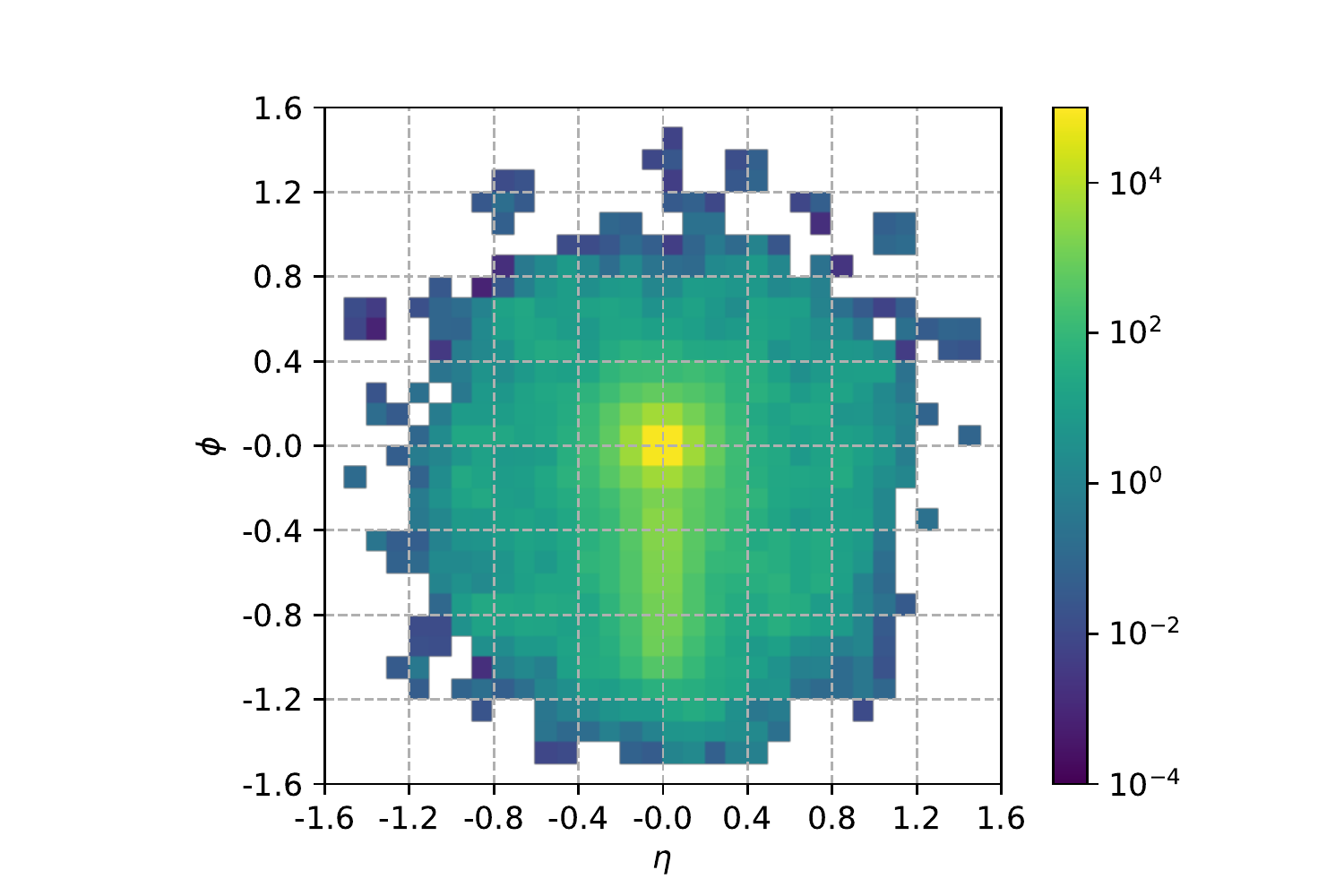}
        \label{1000 background samples}
    }
    \caption{Average jet images of 1000 random samples. The left one is from the signal and the right one is from the background.}
    \label{fig:1000 random samples}
\end{figure}

In this section, we show the attention generated by ABNN1 using different average jet images as queries, so that we can visually understand what the network has learned. First, we randomly select 1000 samples of the signal and background respectively. Fig.\ref{fig:1000 random samples} shows their average jet images.

\begin{figure}[htp]
    \centering
    \subfigure[]{
        \includegraphics[scale=0.5]{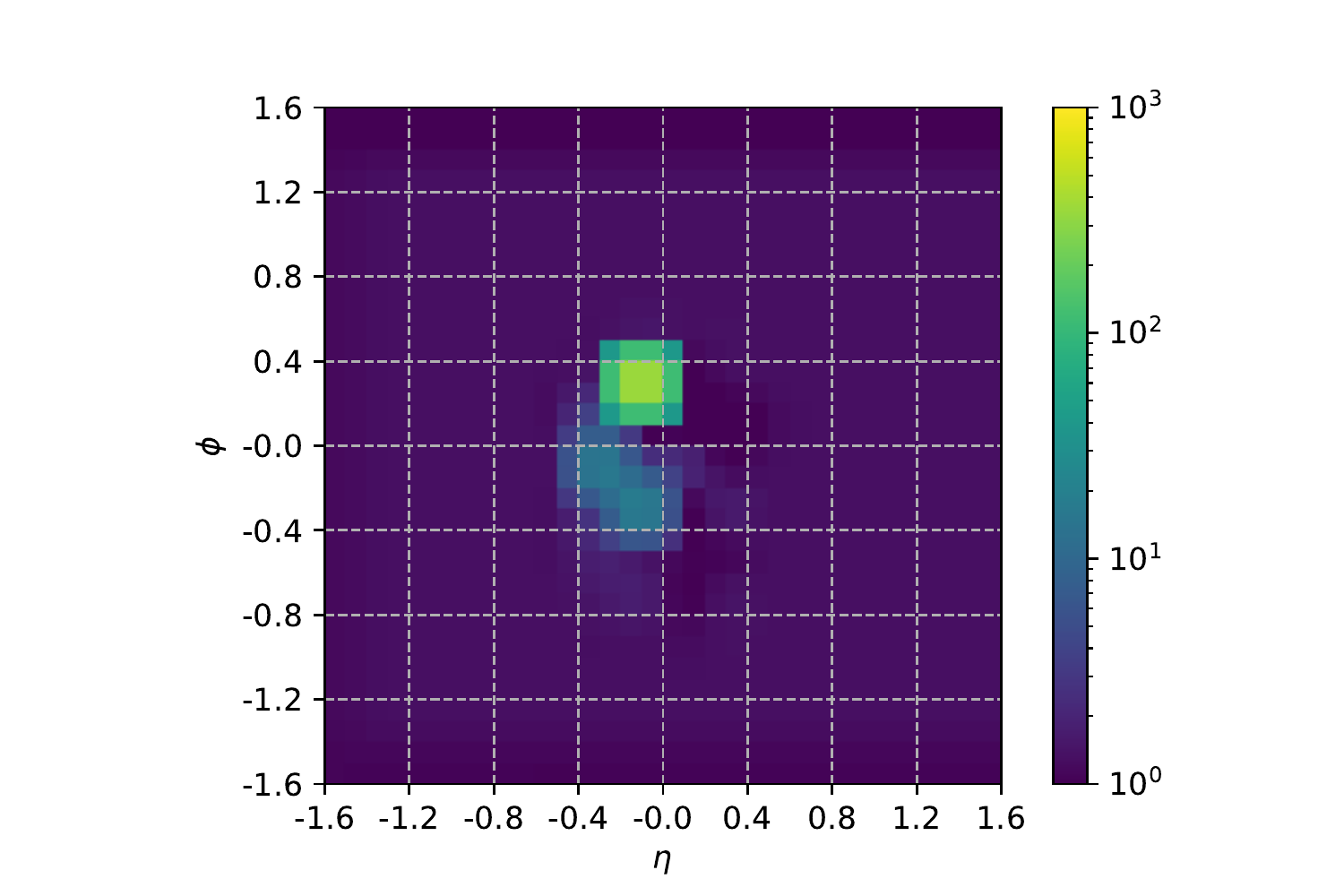}
        \label{subfig: signal attention s}
    }
    \subfigure[]{
        \includegraphics[scale=0.5]{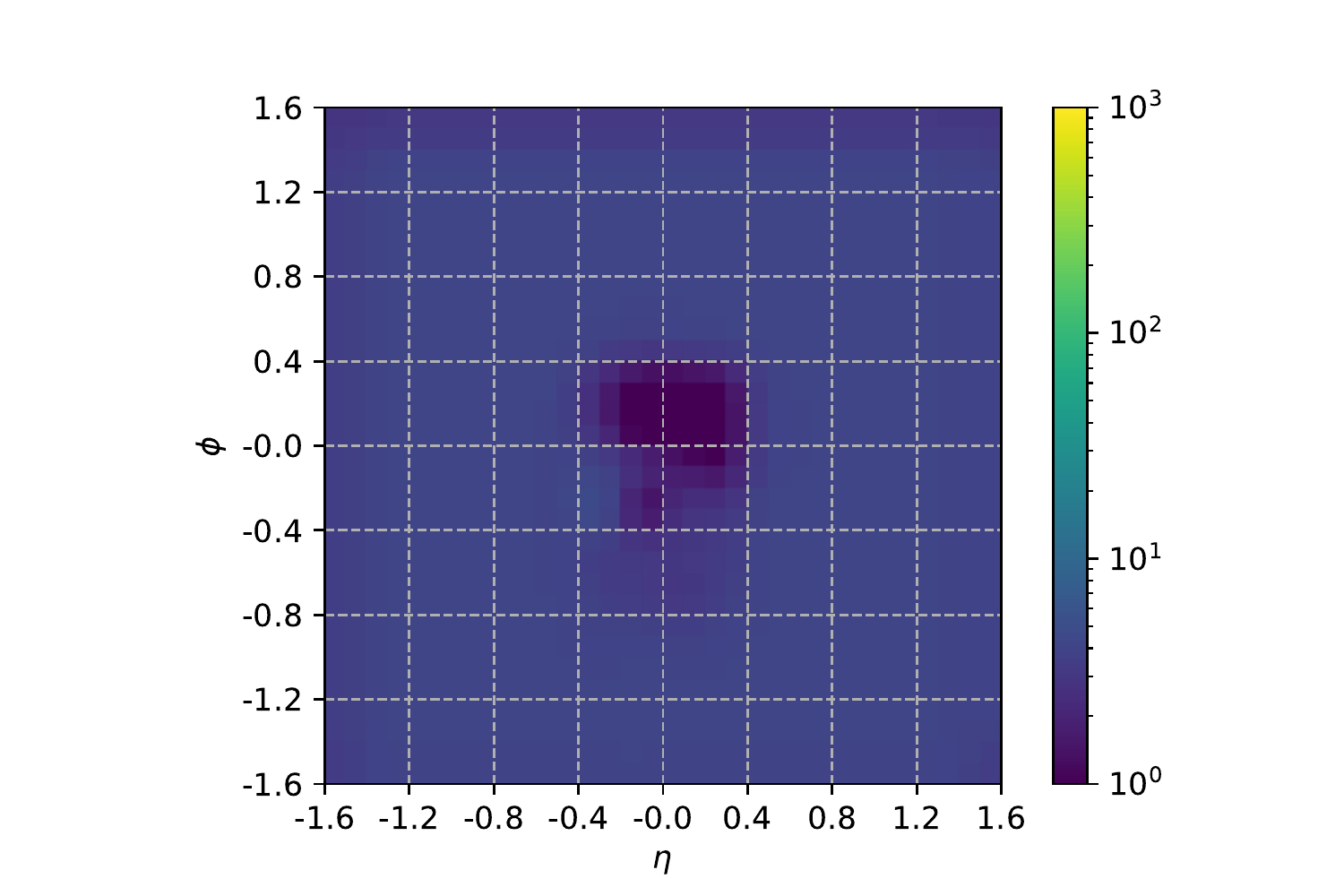}
        \label{subfig: signal attention b}
    }
    \subfigure[]{
        \includegraphics[scale=0.5]{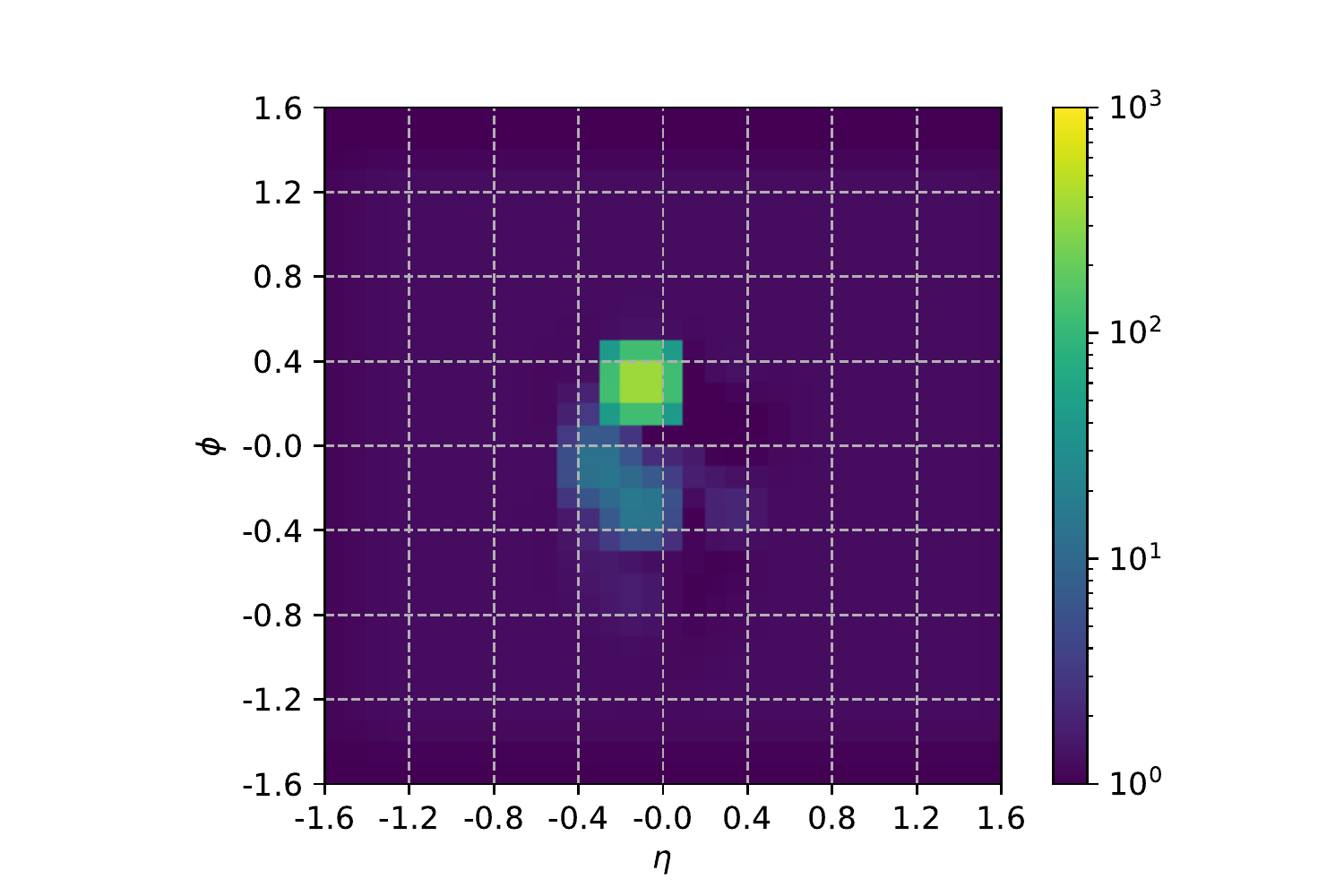}
        \label{subfig: background attention s}
    }
    \subfigure[]{
        \includegraphics[scale=0.5]{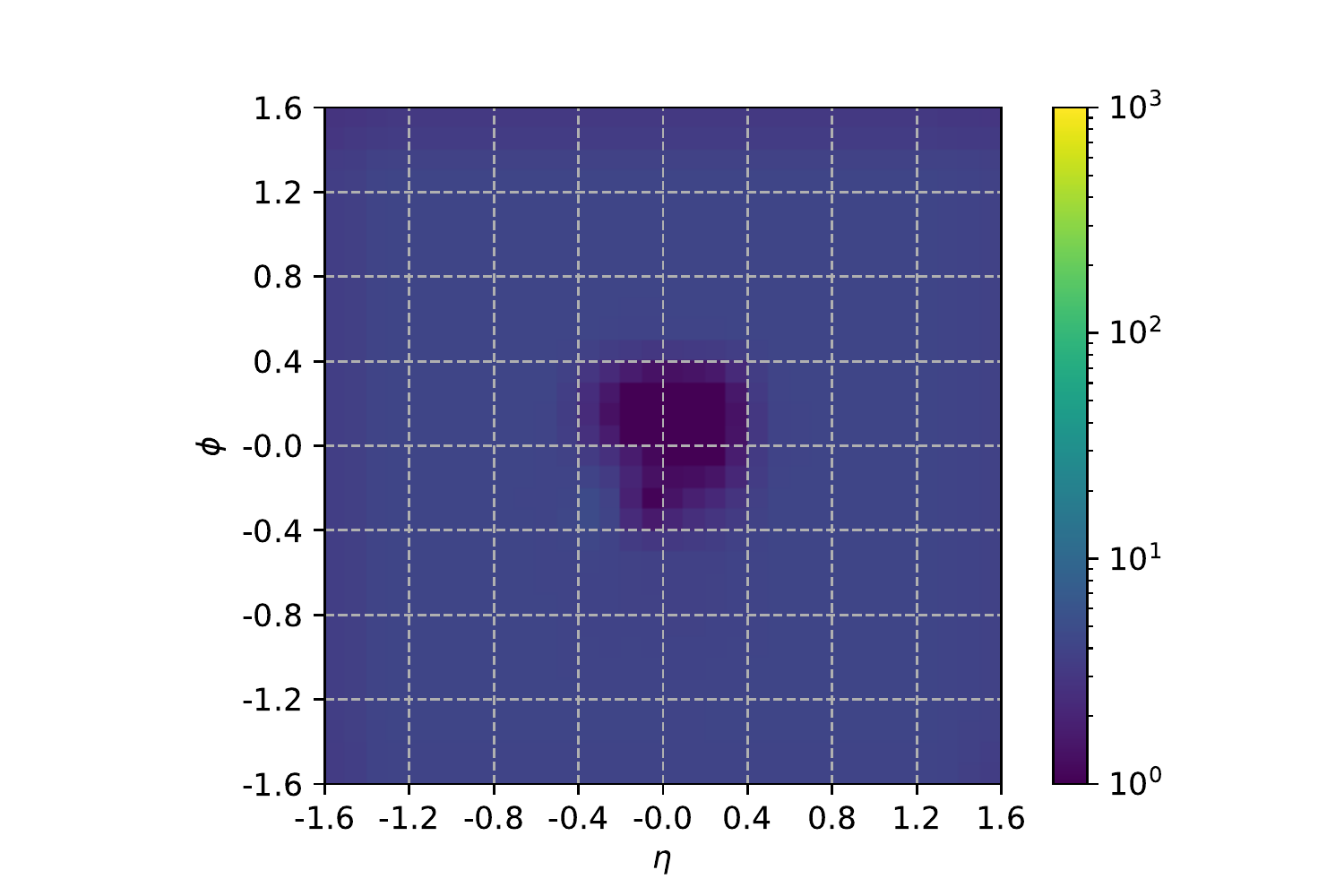}
        \label{subfig: background attention b}
    }
    \caption{Attention. 
             The rows represent the attention obtained by samples from the signal and the background respectively.
             The columns represent the attention by the signal and background average jet images as queries respectively.}
    \label{fig:attention}
\end{figure}

Compared with the background, the subleading subjet in the signal shows stronger spatial characteristics, while in the background, its position is continuous, corresponding to the characteristics of the $\Delta R$ distribution. Passing the samples to the network, we get two kinds of attention: one is to use the signal average jet image as the query, and the other query is the background average jet image. Considering the structure of the ABNN1 itself, we actually get the attention in three different dimensions. For the sake of understanding and display, only the attention with the highest dimension ($16, 16$) that locates at the shallow layer of the network is shown here. The deeper attention is more abstract and has fewer pixels, which is hard to explain. The output attention is normalized into the range $[0, 1]$. Fig.\ref{fig:attention} shows the attention we get.

It clearly indicates that when the average jet image of the signal is used as the query, the network pays more attention to the pixel where the leading and subleading subjets are located. What attracts us is when the query is from the background average jet image, although the background does contain subjets, the network nearly avoids all these areas and focuses on surroundings. Compared to the subjet pixels, the attention of the remaining pixels is far less, which indicates the former weights more when it comes to classification. 

The two sets of attentions, one for the input signal and the other for background, are almost the same. Considering the designing idea of the ABNNs, this shows that the FEs in the ABNNs becomes a unified feature extractor for both categories of average jet images as expected. We conclude that the network focuses on the area where the subjets are  located, while compressing the contribution from the rest of the area. The average jet image from the signal is used as a template to focus on the subjet, and the background average jet image is used to focus on other area. Although their differences exist both in the area where the subjet is and the surrounding pixels, with the help of the attention mechanism the well-trained network is capable of distinguishing these two differences, showing them in different sets of attention.

\section{DISCUSSION AND OUTLOOK}
\label{VI}

In this work, we propose attention-based convolutional neural networks (ABNN) to get insights from the jet tagging problem. With average jet images as additional inputs, independent stacked CNNs are used as feature extractors. Attention is obtained by calculating the compatibility scores of intermediate feature maps from these extractors. 

We apply ABNNs to classify the weak boson Z decaying dijets from general QCD jets to check the performance. Three kinds of architectures are taken into consideration: ABNNs which contain separate feature extractors for jet images and average jet images and utilize attention mechanism, CNNs which have no attention mechanism but still include average jet images, CNNs which just pass jet image through two separate feature extractors. We demonstrate that the ABNNs outperform the other two kinds of CNNs in classification accuracy and background rejection.

The visualization of attention over the original jet image shows clearly the focus of the network during the classification. Different average jet images bring different attentions: the average signal jet image instructs the network to focus on the subjet pixels, while the background one, which contains subjets as well, precisely avoids the area containing subjets. Compared with the average background jet image, attention is more powerful brought by the average signal jet image, which shows that subjets  weight than surrounding hadrons. These results are consistent with what we know when we perform the event reconstructions.

In addition, we benchmark the best ABNN against the standard stacked CNNs. The results show the ABNN is similar with the CNN in which the max channel number of feature maps is 64. Different numbers of jet images to create average images are also explored to see the impact of increase of surrounding pixel intensities. As a result, the ABNNs have almost the same performace. On the contrary, the performance of CNN11 that without the attention mechanism has declined.

Our work can be extended in various ways. Different network structures can be further explored to apply the attention mechanism, such as using the high-level feature maps as queres in \cite{Jetley2018}, or substituting attention for the convolution operations in \cite{Bello2019}. It is not only applicable to  the jet image but also other representations. As \cite{Qu2019} mentioned, jet image is not an efficient representation due to its sparsity. It seems to have a big potential to gain improvement with attention mechanism applied to other jet representations. More possibilities are left for future works.

\begin{acknowledgments}
Hao Sun is supported by the National Natural Science Foundation of China (Grant No.11675033).
\end{acknowledgments}

\bibliography{jet_image}
% \bibliography{attention}
%\begin{thebibliography}{99}
%\end{thebibliography}
\end{document}